\documentclass[twocolumn,twocolappendix,numberedappendix]{openjournal}

\usepackage{graphicx}
\usepackage{amsmath}
\usepackage{gensymb}
\usepackage{xspace}
\usepackage[dvipsnames]{xcolor}
\usepackage{fontawesome}
\usepackage{bm}
\usepackage{ifthen}
\usepackage{booktabs}
\usepackage{import}
\usepackage{hyperref}

\renewcommand{\equationautorefname}{Eq.}
\renewcommand{\figureautorefname}{Fig.}

\hypersetup{colorlinks=true, allcolors=Blue}
\makeatletter
\let\oldtheequation\theequation
\renewcommand\tagform@[1]{\maketag@@@{\ignorespaces#1\unskip\@@italiccorr}}
\renewcommand\theequation{(\oldtheequation)}
\makeatother

\newcommand{\Autoref}[1]{%
  \begingroup
  \def\sectionautorefname{Section}%
  \def\subsectionautorefname{Subsection}%
  \def\chapterautorefname{Chapter}%
  \def\figureautorefname{Figure}%
  \def\equationautorefname{Equation}%
  \def\tableautorefname{Table}%
  \autoref{#1}%
  \endgroup
}

\newcommand\rd{{\rm d}}
\newcommand{\ksmpc}{$\,\mathrm{km}\,\mathrm{s}^{-1}\,\mathrm{Mpc}^{-1}$}
\newcommand\balpha{\bm{\alpha}}
\newcommand\btheta{\bm{\theta}}

\newcommand\bxi{\bm{\xi}}
\newcommand\bx{\bm{x}}

\newcommand\bet[1]{\mathcolor{Maroon}{#1}}

\newcommand\vect[1]{\boldsymbol{#1}}

\newcommand{\uvect}[1]{\hat{\vect{#1}}}

\newcommand\e[1]{_{\mathrm{#1}}}
\newcommand\h[1]{^{\mathrm{#1}}}

\newcommand{\delimiters}[4][]{
\ifthenelse{ \equal{#1}{1} }{  #2 #3 #4  }
					{ \ifthenelse{\equal{#1}{2}}{ \big#2 #3 \big#4 }
						{ \ifthenelse{\equal{#1}{3}}{ \Big#2 #3 \Big#4 }
							{ \ifthenelse{\equal{#1}{4}}{ \bigg#2 #3 \bigg#4 }
								{ \ifthenelse{\equal{#1}{5}}{ \Bigg#2 #3 \Bigg#4 }
									{ \left#2 #3 \right#4 }
								}
							}
						}
					}
													}
													
\newcommand{\pa}[2][]{\delimiters[#1]{(}{#2}{)}}
\newcommand{\pac}[2][]{\delimiters[#1]{[}{#2}{]}}
\newcommand{\paac}[2][]{\delimiters[#1]{\{}{#2}{\}}}

\definecolor{blue4}{RGB}{0,0,143}
\definecolor{red4}{RGB}{143,0,0}
\definecolor{orange}{RGB}{255,128,0}
\definecolor{darkcyan}{RGB}{0,128,128}
\definecolor{olive}{RGB}{0,128,0}
\definecolor{purple}{RGB}{128,0,128}
\definecolor{cyan2}{RGB}{0,255,255}
\definecolor{fushia}{RGB}{255,0,255}
\definecolor{mygray}{gray}{0.5}
\definecolor{lightgray}{gray}{0.85}

\newcommand{\zone}{$z_1 = 0.1840$\xspace}
\newcommand{\ztwo}{$z_2 = 1.8845$\xspace}
\newcommand{\zs}{$z\e{s}= 2.382$\xspace}

\begin{document}
\title{Lessons from double-plane lenses for galaxy profiles, the Hubble tension, and the mass-sheet degeneracy}

\author{Daniel P. Johnson$^{1 \ast}$, 
Martin Millon$^{2,3}$, 
Pierre Fleury$^1$, 
Giacomo Queirolo$^1$, 
Tian Li$^4$, 
Wolfgang J. R. Enzi$^5$, 
Daniel J. Ballard$^6$,
Duncan Bowden$^7$,
Thomas E. Collett$^4$,
Daniel Gilman$^8$}
\thanks{$^\ast$daniel.johnson@umontpellier.fr}

\affiliation{$^1$ Laboratoire Univers et Particules de Montpellier, CNRS \& Université de Montpellier,\\Parvis Alexander Grothendieck, Montpellier, France, 34090}
\affiliation{$^2$ Institute for Particle Physics and Astrophysics, ETH Zurich,
Wolfgang-Pauli-Strasse 27, CH-8093 Zurich, Switzerland}
\affiliation{$^3$ D\'epartement de Physique Th\'eorique, Universit\'e de Gen\`eve, 24 quai Ernest-Ansermet, CH-1211 Gen\`eve 4, Switzerland}
\affiliation{$^4$  Institute of Cosmology and Gravitation, University of Portsmouth, Burnaby Road, Portsmouth, PO1 3FX, UK}
\affiliation{$^5$ INAF $-$ Istituto di Radioastronomia, Via Gobetti 101, I$-$40129, Bologna, Italy}
\affiliation{$^6$ Sydney Institute for Astronomy, School of Physics, A28, The University of Sydney, NSW 2006, Australia
}
\affiliation{$^7$ Department of Physics, University of Oxford, Keble Road, Oxford OX1 3RH, UK}
\affiliation{$^8$  Kavli Institute for Cosmological Physics, University of Chicago, Chicago, IL 60637, USA}

\begin{abstract}
Double-plane lenses, like single-plane lenses, are subject to the mass-sheet degeneracy, a fundamental limit to our knowledge of the deflector mass profiles from lensing. However, with a prior on the relative expansion history, unlike in the single-plane case, this degeneracy can be constrained with the modelling of lens images, without the need for time delays, velocity dispersion measurements or a prior on the Hubble constant $H_0$. 

We show that double-plane lenses are sensitive to a specific combination of the mass-sheet components of the two lens planes. Using literature values of the cosmological scaling factor from the double-plane lenses J0946 and AGEL0353, a measurement from the point-source images of J1721 (the ``Einstein zig-zag'' lens), and a reference value of $\Omega\e{m}$, we show the constraints that can be placed on the mass-sheet component of each of these systems. We show that the additional information brought by images of a second source can improve measurements of $H_0$ from individual time-delay lenses by up to $\sim50\%$, depending on certain key assumptions. Using forecasted \textit{Euclid} and LSST double-plane lens samples, we show that these samples will enable strong constraints on the population of mass sheets in lens galaxies. Such population constraints will have implications for a range of strong-lensing science applications, in particular in constraining $H_0$, dark energy, and the dark matter profiles of galaxies.

\end{abstract}

\begin{keywords}
    {Strong gravitational lensing, cosmology}
\end{keywords}

\section{Introduction}
\label{sec:introduction}
%
When a sufficiently massive and dense galaxy or cluster lies on or close to the line of sight between our telescopes and a distant source of light, its light may be strongly lensed, such that we observe arcs, multiple images or Einstein rings. The path of this lensed light carries information about the dominant lensing mass, the distribution of mass along the line of sight but external to this main lens, and the background cosmology through which it propagates, and if these are carefully modelled, they can be constrained with the images we observe.

The power of strong gravitational lensing as a cosmological probe is greatly enhanced by two special classes of lensed systems, time-delay and double-plane lenses. When a bright, compact object such as a quasar or supernova, whose luminosity may fluctuate on relatively short timescales, is multiply imaged by a strong lens, the light curves of these multiple images can be compared, and the relative differences in arrival time between these trajectories can be measured. These relative time delays are proportional to a ratio of angular diameter distances such that, if the lens mass is well-constrained, these systems can be used to measure the Hubble constant $H_0$ \citep{Refsdal_1964,Suyu_2010,Birrer_2022}. For individual lens galaxies, this has achieved precisions up to $6.5\%$ \citep[e.g.][]{sheu_2026}, improving to 3\% when combining a sample of 11 systems \citep{TDCOSMO_2026}. 

On rare occasions, the same lens galaxy may strongly lens two background sources offset from one another in redshift, such that the most distant source may also be lensed by the foreground source/background lens. These ``double-plane lens'' (DPL) systems enable a measurement of the  ``cosmological scaling factor'' $\eta$, a combination of angular diameter distances which does not depend on $H_0$, but is instead sensitive to the relative expansion history of the universe \citep{Collett_2012,Sharma_2023}. Such systems have been used to place constraints on $\Omega\e{m}$ and the dark energy equation of state, $w(z)$ \citep{Collett_2014,Sahu_2025,Bowden_2025,ballard_2026}, and in detailed studies of the lens mass profile \citep[e.g.][]{Li_2026}. The number of known DPLs has grown steadily in the last decade \citep{Gavazzi_2008,Tu_2009,Tanaka_2016,Schuldt_2019,Lemon_2023,Bolamperti_2023,barone_2025,Li25}, and is expected to rise dramatically with the advent of stage-IV surveys such as \textit{Euclid} and LSST \citep{Euclid_2022,LSST_2021,Sharma_2022,Li25}.

The state of modern cosmology is such that independent constraints on $H_0$ and $w(z)$ have never been more important. So-called ``early universe'' probes from BAO and the CMB, coupled with a flat-$\Lambda$CDM cosmological model, predict values of $H_0$ which are consistently lower than those derived from most ``late-universe'' distance ladder methods. Comparing the local distance network result from \cite{Casertano_2026} with the early universe result from \cite{Camphuis_2026}, this tension stands at $7.1\sigma$, with no straightforward observational or theoretical explanation at hand \cite[see e.g.][]{Cai_2026,schöneberg_2026}. Additionally, recent results from combined BAO, CMB and supernova analyses can be interpreted as evidence for dynamical dark energy, i.e. a $w(z)$ incompatible with a cosmological constant at $\sim3\sigma$ \citep{DES_2026}. The appeal of time-delay and DPL science as independent probes of these phenomena is therefore greater than ever \citep{Birrer_2022,Sharma_2023}.

The key source of uncertainty in both of these methods is the so-called ``mass-sheet degeneracy'' \citep[MSD,][]{Falco_1985}. The mass model of a given lens is underdetermined by lensed images and time delays alone, even when multiple images are present, without a fixed cosmological model. Any change to the modelled projected mass which mimics a change to the cosmology will inevitably be degenerate with the inferred cosmological parameters. Put differently, if the modelled lens profile is a rescaling of the true profile by a factor $\lambda$, and a corresponding constant uniform mass sheet is added to the model, lensed image positions and time delay ratios remain unchanged, but model-predicted relative time delays are rescaled. When these relative time delays are then measured, the inferred Hubble constant is biased by a factor $\lambda$ \citep[see][and others]{Gorenstein_1988,Saha_2000,Wucknitz_2002,Liesenborgs_2012,Schneider_2013,Birrer_2016,Gomer_2020,Birrer_2022}. 

An MSD-driven mismodelling of the lens mass can originate from two parts of the model. A uniform projected over or underdensity along the line of sight to the lens and source is umeasurable, and hence is usually constrained using the ``weighted number counts'' technique, by which the distribution of galaxies in the lens environment is compared to simulations in which this projected density contrast can be estimated \citep[e.g.][]{Suyu_2010,Rusu_2017,Wells_2023,johnson_2026b}.

The ``internal MSD'' refers to the fact that a mass-sheet transformation of the main lens itself is similarly undetectable without independent constraints \citep{Schneider_2013}. Lens galaxies are typically modelled with simple parametric models, such as elliptical power laws \citep{Tessore_2015} or a dark-matter Navarro--Frenk--White plus baryonic Sérsic profile. Simulations are not yet able to resolve baryonic effects on the necessary scales to unambiguously predict lens model profiles, and it has also been shown that a change of dark matter model could mimic a change in $\lambda$ \citep{Blum_2020,Blum_2021}. The internal mass-sheet component of lens galaxies is typically constrained with velocity dispersion measurements \citep[see e.g.][]{Treu_2002,Suyu_2013,sheu_2026}. 

It is sometimes argued that the images of multiple sources at different redshifts lensed by the same object can be used to break the MSD. Unfortunately, this is only true if the cosmology is fixed, and the mismodelled mass is located in the first (foreground) lens plane only. If these assumptions are relaxed, the degeneracy remains present, and affects both the determination of $H_0$ and constraints on the relative expansion history from $\eta$ \citep{Liesenborgs_2008,Schneider_2014,Schneider_2019,Teodori_2022,Johnson_2025,teodori_2026,Johnson_2026}.

The sensitivity of $\eta$ measurements to the MSD could be seen as a curse for constraining the relative expansion history, but a blessing for studying the mass-sheet degeneracy and the mass profiles of galaxies. If a prior is placed on cosmological parameters such as $\Omega\e{m}$ and $w$, and the external component of the MSD is estimated from simulations and weighted number count techniques, the freedom of the internal components in the two planes of a DPL is greatly restricted, and it becomes possible to study the Hubble tension and galaxy profiles (and hence dark matter models) in a way which would be impossible from single-plane lenses alone. 

The connection between the MSD and the relative expansion history in double-plane lenses is the main motivation of this work, which demonstrates that these systems can provide unique insights for galaxy mass profiles and strong lensing cosmography. In \autoref{sec:theory}, we review the theoretical aspects of double-plane and time-delay lenses. In \autoref{sec:MSD}, we recap the single and double-plane mass-sheet degeneracies, and their impact on our inference on cosmological parameters. In \autoref{sec:mass-sheet constraints}, we discuss the physical implications of an internal mass-sheet transformation, and then detail the novel constraints which can be made from DPLs, and in particular when time delays are also present. In \autoref{sec:zigzag_application}, we apply these techniques to the ``Einstein zig-zag lens'', J1721+88842, and obtain constraints on mass profiles of both lens planes. In \autoref{sec:literature_application}, we consider the DPLs J0946+1006 and AGEL035346$-$170639, and show that these systems enable even tighter constraints. In \autoref{sec:forecasts}, we anticipate the constraining power of future samples of systems observed with surveys such as \textit{Euclid} and LSST, and argue that these will enable unprecedented population-level constraints on the internal mass-sheet components of strong lenses. Finally, in \autoref{sec:forecasts_time_delays}, we make tentative comments on the $H_0$ constraining power of DPLs with time delays. We discuss the implications of this work in \autoref{sec:discussion}. The key parameters used throughout this work, and their interpretation, are listed in \autoref{tab:parameters}.

\section{Double-plane lensing}
\label{sec:theory}

\begin{figure*}
\centering
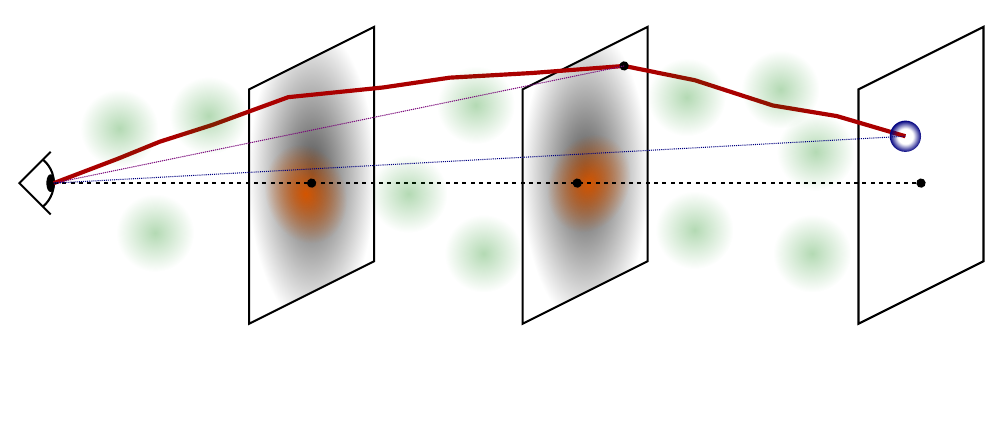
\caption{The geometry of double-plane lensing with line-of-sight perturbers (shown in \textcolor{olive}{green}) and internal mass-sheet components (shown in \textcolor{gray}{gray}). Light originating from a source in plane 3 experiences additional (de)focusing by the LOS perturbers, which we model as pure convergence for the purposes of this paper. In plane 1, and possibly in plane 2 as well, it is strongly lensed. \vspace{1em}}
\label{fig:double-plane_lensing}
\end{figure*}

\begin{table}
\centering
\begin{tabular}{p{0.27\columnwidth}p{0.69\columnwidth}}
\hline
Quantity & Definition \\
\hline
$\bxi_l$ & transverse separation vector in the $l^\mathrm{th}$ plane \\
$D_{ml}$ &  background angular diameter distance to plane $l$ as seen from plane $m$, \autoref{eq:D_ml}  \\
$\btheta_l$ & angular position defined by $\bxi=D_l\btheta_l$, \autoref{eq:multi-plane_lens} \\
$\kappa_{ml}$ & weak lensing convergence to $l$ as seen from $m$, \autoref{eq:kappa_ij_delta} \\
$\bm{\alpha}_l(\bm{\theta})$ & effective displacement angle due to lens $l$ (including weak lensing corrections), \autoref{eq:alpha1_los} and \ref{eq:alpha2_los} \\
$\psi_l(\bm{\theta})$ & effective lensing potential of lens $l$ (including weak lensing corrections), \autoref{eq:alpha1_los} and \ref{eq:alpha2_los} \\
$\eta$ & effective cosmological scaling factor (including weak lensing corrections), \autoref{eq:los_eta} \\
$\bar{\eta}$ &  cosmological scaling factor in the background cosmology, \autoref{eq:kappa_eta_and_bar_eta} \\
$\kappa_\eta$ &  LOS  correction to the cosmological scaling factor \autoref{eq:kappa_eta_and_bar_eta} \\
$\kappa\h{lens}_{l}$ & total convergence in plane $l$ (including the LOS  contribution) \autoref{eq:kappa1_kappa2}  \\
$\phi_{ml}\left(\bm{\theta}_m,\bm{\theta}_l\right)$ & Fermat potential for a source at $l$ lensed by a lens at $m$ (including weak lensing corrections), \autoref{eq:phis} \\
$\tau_{ml}$ & time delay scale (including weak lensing corrections), \autoref{eq:tau_ml_los} \\
$\kappa^\mathrm{LOS}_{ml}$ & LOS  component of $\tau_{ml}$, \autoref{eq:kappa_LOS} \\
$\tilde{\kappa}_l$ & mass-sheet transformed mass profile in plane~$l$, \autoref{eq:mst_1_k} and \ref{eq:mst_2_k} \\
$\lambda$ & mass-sheet parameter \\
$\nu_3$ & ratio of the mass-sheet transformed to true source position, \autoref{eq:r_eta_and_nu_3} \\
$r_\eta$ & ratio of the mass-sheet transformed to true cosmological scaling factors \autoref{eq:r_eta_and_nu_3} \\
$\tilde{\tau}_{ml}$ & mass-sheet transformed time-delay scale,  \\
$\tilde{\eta}$ & measured (and potentially mass-sheet transformed) cosmological scaling factor  \\
$H_0$ & true Hubble constant \\
$\tilde{H}_0$ & measured (and potentially mass-sheet transformed) Hubble constant, \autoref{eq:H0_msd} \\
$\kappa\e{mod}(\btheta)$ & fiducial mass profile used in the lens modelling (typically an EPL) \\
$\zeta_l$ & mass sheet convergence in plane $l$, including LOS corrections, \autoref{eq:fid+core} \\
$\kappa\e{c}(\btheta)$ & functional form of the approximate mass-sheet ($\approx 1$ on lensing scales), \autoref{eq:fid+core} \\
$\zeta_{12}$ & a combination of $\zeta_1$ and $\zeta_2$ constrained by comparing $\tilde{\eta}$ and $\eta$, \autoref{eq:zeta_12_definition} \\
$V_l$ & stellar kinematics constraint in plane $l$, \autoref{eq:v1} and \ref{eq:v2} \\
$\bar{\zeta}_l$ & convergence of the mass sheet component in plane $l$, excluding LOS corrections, \autoref{eq:zeta1_with_los} and \ref{eq:zeta2_with_los} \\
\hline 
\end{tabular}
\caption{Key parameters used in this work, and the quantities they correspond to.}
\label{tab:parameters}
\end{table}
%
\subsection{Multi-plane lensing}
%
Suppose light originating at a source s is seen by an observer o after passing through $N$ thin matter planes. As it passes through the $l$-th plane with surface mass density $\Sigma_l(\vect{\xi})$, light is deflected by an angle
\begin{equation}
    \uvect{\alpha}_l(\vect{\xi}) = \frac{4G}{c^2}\int\rd^2 \bx' \; \Sigma_l(\vect{\xi}')\frac{\vect{\xi}-\vect{\xi}'}{|\vect{\xi}-\vect{\xi}'|^2},
    \label{eq:deflection_angle}
\end{equation}
where $\vect{\xi}=D_l\btheta_l$ is the physical transverse separation vector in the $l$-th plane, related to the unlensed angular position $\btheta_l$ (at which the light would be observed in the absence of the planes between o and $l$) via the (background) angular diameter distance $D_l$ to plane $l$. These angular positions $\btheta_l$ are determined recursively from positions $\btheta_m$ and deflection angles $\uvect{\alpha}_m$, where $m<l \leq N$, via the multi-plane lens recursion,
\begin{equation}
    \btheta_l = \btheta - \sum^{l-1}_{m=1}\frac{(1-\kappa_{ml})}{(1-\kappa_l)}\frac{D_{ml}}{D_{l}}\hat{\bm{\alpha}}_m(\btheta_m),
    \label{eq:multi-plane_lens}
\end{equation}
where $\btheta = \btheta_1$ is the angular position at which the light is observed, and $D_{ml}$ refers to the angular diameter distance to $l$ as seen from $m$ in the background cosmology, which, under the assumption of no spatial curvature, is given by
\begin{equation}
    D_{ml}=\frac{\chi_l-\chi_m}{1+z_l},
    \label{eq:D_ml}
\end{equation}
where $z_l$ is the cosmological redshift of $l$ and $\chi_l, \chi_m$ are the comoving distances from the observer to $l$ and $m$. $\kappa_{ml}$ is the corresponding line-of-sight convergence to plane~$l$ as seen from plane $m$, and $\kappa_l$ is the convergence to plane~$l$ as seen from the observer plane. From \cite{Fleury_2019a}, assuming a spatially flat universe, $\kappa_{ml}$ can be written as a weighted integral of the matter density contrast $\delta\e{m}$, 
\begin{equation}
    \kappa_{ml} = \frac{3}{2}\Omega\e{m}H_0^2\int^{\chi_l}_{\chi_m}\rd \chi \; (1+z)K_{ml}(\chi)\delta\e{m}[\eta_0-\chi,\chi,\chi\btheta],
    \label{eq:kappa_ij_delta}
\end{equation}
where $\Omega\e{m}$ is the matter density parameter today, $\eta$ is the conformal time, $\eta_0$ is its value today, and the integration kernel is given by
\begin{equation}
    K_{ml}(\chi) \equiv \frac{(\chi_l-\chi)(\chi-\chi_m)}{\chi_l-\chi_m}. \label{eq:weight_function}
\end{equation}
%
\subsection{Double-plane lensing and the scaling factor}
%
A double-plane lens system, as depicted in \autoref{fig:double-plane_lensing}, consists of a source at $z_3$ lensed by a background lens at $z_2$ and a foreground lens at $z_1$, with the light from lens 2 typically lensed by the foreground lens as well. For such a system, we can write
\begin{align}
    \btheta_2 &= \btheta - \bm{\alpha}_{1}(\btheta), \label{eq:eff_lens_1} \\
    \btheta_3 &= \btheta - \eta\,\bm{\alpha}_1(\btheta)-\bm{\alpha}_2(\btheta_2). \label{eq:eff_lens_2}
\end{align}
In these expressions, as in \cite{Johnson_2026}, quantities are defined to absorb the contribution from the line-of-sight convergence, thus the displacement angles~$\vect{\alpha}_l$ (defined under the $\eta$ convention for sources in the plane $l+1$, see \autoref{sec:beta_formalism} for an alternative) are given by 
\begin{align}
    \bm{\alpha}_1(\btheta)  &\equiv \frac{\rd \psi_1(\btheta)}{\rd\btheta} = \frac{1-\kappa_{12}}{1-\kappa_2} \frac{D_{12}}{D_2}\uvect{\alpha}_1[(1-\kappa_1)D_1\btheta], \label{eq:alpha1_los} \\ 
    \bm{\alpha}_2(\btheta_2) &\equiv \frac{\rd \psi_2(\btheta_2)}{\rd\btheta_2} = \frac{1-\kappa_{23}}{1-\kappa_3}\frac{D_{23}}{D_3}\uvect{\alpha}_2[(1-\kappa_2)D_2\btheta_2], \label{eq:alpha2_los}
\end{align}
where $\psi_{l}(\btheta_l)$ is the lensing potential of plane $l$, once again defined to absorb the corresponding line-of-sight convergence terms. The cosmological scaling factor $\eta$ is
\begin{equation}
    \eta \equiv \frac{(1-\kappa_2)(1-\kappa_{13})}{(1-\kappa_{12})(1-\kappa_3)}\frac{D_2D_{13}}{D_{12}D_3} = \pa{1-\kappa_\eta}\bar{\eta}, \label{eq:los_eta}
\end{equation}
with
\begin{equation}
    1-\kappa_\eta = \frac{(1-\kappa_2)(1-\kappa_{13})}{(1-\kappa_{12})(1-\kappa_3)}, \qquad \bar{\eta} = \frac{D_2D_{13}}{D_{12}D_3}.
    \label{eq:kappa_eta_and_bar_eta}
\end{equation}

Associated with a displacement angle $\balpha_l(\btheta)$ is the convergence $\kappa\h{lens}_l(\btheta)$, defined as the projected surface mass density $\Sigma_l(\btheta)$ in that plane in units of the critical density~$\Sigma\h{crit}_l$,
\begin{equation}
    \Sigma\h{crit}_l = \frac{1}{4\pi G} \frac{1-\kappa_{l+1}}{(1-\kappa_{l})(1-\kappa_{l(l+1)})}\frac{D_{l+1}}{D_{l}D_{l(l+1)}},
\end{equation}
in which the parameter definition once again includes the line-of-sight contribution, such that
\begin{equation}
    \kappa\h{lens}_l(\btheta_l) \equiv \frac{\Sigma_l(\btheta_l)}{\Sigma\h{crit}_l} = \frac{1}{2}\nabla^2\psi_l(\btheta_l), \label{eq:kappa1_kappa2} 
\end{equation}
for $l\in\{1,2\}$. Note the distinction between the \textit{total} convergence $\kappa\h{lens}_l$ in plane $l$ (with line-of-sight contributions), and the weak lensing convergence $\kappa_l$ which acts on a source in plane $l$ as seen from the observer.
\subsection{Time delay cosmography}
From e.g. \cite{Fleury_2021b}, the light travel time from $\btheta_3$ via $\btheta_2$ and $\btheta$ to the observer is given by 
\begin{equation}
    T(\btheta,\btheta_2,\btheta_3) = \tau_{12} \, \phi_{12}\pa[1]{\btheta,\btheta_2} + \tau_{23} \, \phi_{23}\pa[1]{\btheta_2,\btheta_3}.
    \label{eq:time_delays}
\end{equation}
$\phi_{12}\pa[1]{\btheta,\btheta_2}$ and $\phi_{23}\pa[1]{\btheta_2,\btheta_3}$ are Fermat potentials, dimensionless quantities given by
\begin{equation}
    \phi_{ml}(\btheta_m,\btheta_l)
    = \frac{1}{2}(\btheta_m-\btheta_l)^2-\psi_m\pa[1]{\btheta_m},
    \label{eq:phis}
\end{equation}
for $m,l\in\{1,2,3\}$. $\tau_{12}$ and $\tau_{23}$ are the time-delay scales, a combination of background angular diameter distances and convergence factors,\footnote{In the literature, $\tau_{ml}$ (or $D^{\Delta t}_{ml}=c \, \tau_{ml}$, the time-delay distance) is often used to refer specifically to the background quantity, with no line-of-sight convergence terms. Following the previous conventions in this paper and in \cite{Johnson_2026}, we find it convenient to absorb these contributions into the definition of $\tau_{ml}$.}
\begin{gather}
    \tau_{ml} = (1+z_m)\pa{1-\kappa\h{LOS}_{ml}}\frac{D_mD_l}{D_{ml}}, \label{eq:tau_ml_los} 
\end{gather}
where
\begin{equation}
    1-\kappa\h{LOS}_{ml} \equiv \frac{(1-\kappa_m)(1-\kappa_l)}{1-\kappa_
    {ml}}. \label{eq:kappa_LOS}
\end{equation}
The background angular diameter distances appearing in Eqs.~\ref{eq:multi-plane_lens} and \ref{eq:los_eta} can be expressed as an integral of the dimensionless expansion function~$H(z)/H_0$ (whence the dependence on cosmological parameters such as $\Omega\e{m}$), and with a prefactor proportional to $H_0^{-1}$.
However, these distances only ever appear in ratios, such that the dependence on $H_0$ cancels. In theory, this inobservability is broken by measurements of the relative time delays between the multiple images of a strongly lensed source, as the $H_0$ dependence is reintroduced by the additional distance factors in \autoref{eq:tau_ml_los}. Thus, if these relative time delays and the redshifts of the lens and source planes are measured accurately, then constraints on the lensing potentials allow constraints on $H_0$. In practice, however, such a constraint is fundamentally limited by the mass-sheet degeneracy.
%
\section{The mass-sheet degeneracy}
\label{sec:MSD}
%
The mass-sheet degeneracy is the most famous and amongst the most limiting of the known systematics in strong lensing. In this section, we recap the single and double-plane MSDs, and their effect on $H_0$ and $\eta$ constraints from lensing.
%
\subsection{A recap of the single-plane mass-sheet degeneracy}
%
Consider first the single-plane case, where light is observed at $\btheta$ after lensing by a mass profile $\kappa(\btheta)$. It was shown in \cite{Falco_1985} that there exists an infinite family of mass models $\tilde{\kappa}$,
\begin{equation}
    \tilde{\kappa}(\btheta) = \lambda\kappa(\btheta) + (1-\lambda)
    \label{eq:single_plane_msd}
\end{equation}
which can perfectly reproduce the observed image positions and relative magnifications. 

Physically, this transformation rescales $\kappa(\btheta)$ by the mass-sheet parameter $\lambda$, while simultaneously adding an infinite sheet with homogeneous convergence $1-\lambda$. It affects the inferred source position, but because this is fundamentally unobservable, the degeneracy persists \citep{Schneider_2014b}. It has been shown that this degeneracy is not fully broken by the presence of multiple sources at different redshifts \citep{Schneider_2019}, nor by measurements of time delay ratios \citep{Falco_1985,Kochanek_2004}. While knowing the absolute magnification of the source would in theory allow it to be broken, lensed standard candles are extremely rare, and complications from microlensing means that some uncertainty persists \citep[see][]{yahalomi_2017,weisenbach_2024,Arendse_2025}. 

While the transformation of $\kappa(\btheta)$ by some $\lambda$ leaves image positions and relative magnifications unchanged, the relative time delay between images is rescaled by~$\lambda$. However, as time delays are also proportional to $H_0^{-1}$, lensing observables can only constrain the combination~$\lambda H_0$.\footnote{This is often instead expressed as $\lambda^{-1}H_0$, where $\lambda$ is taken to represent the physical mass-sheet in the first plane. We find this form less convenient for DPLs, though the two are equivalent.} Extracting $H_0$ from $T(\btheta)$ requires independent constraints on the lens mass, and conversely extracting $\lambda$ from $T(\btheta)$ requires a sufficiently tight prior on $H_0$. 
%
\subsection{The double-plane mass sheet degeneracy}
\label{sec:double_plane_mst}
%
In the double-plane case, images of the first source/second deflector constrain $\kappa\h{lens}_1(\btheta)$, while images of the second source constrain $\kappa\h{lens}_1(\btheta)$, $\kappa\h{lens}_2(\btheta_2)$ and $\eta$. \cite{Schneider_2014,Schneider_2019} show that these constraints do not alleviate the mass-sheet degeneracy as it applies to $\kappa\h{lens}_1(\btheta)$, and $\kappa\h{lens}_2(\btheta_2)$ is subject to a similar degeneracy. Image positions and magnification ratios are perfectly reproduced by transformed mass profiles taking the form 
\begin{align}
    \tilde{\kappa}\h{lens}_1(\btheta) &= \lambda\kappa\h{lens}_1(\btheta)+\pa{1-\lambda}, \label{eq:mst_1_k} \\
    \tilde{\kappa}\h{lens}_2(\tilde{\btheta}_2) &= \nu_3\lambda^{-1}\kappa\h{lens}_2\pa{\lambda^{-1}\tilde{\btheta}_2} \nonumber \\
    &\quad
    + \tilde{\eta}\pa{1-r_\eta\nu_3\lambda^{-1}}, \label{eq:mst_2_k}
\end{align}
with the corresponding transformed lensing potentials
\begin{align}
    \tilde{\psi}_1(\btheta) &= \lambda\psi_1(\btheta)+\frac{1}{2}\pa{1-\lambda}\btheta^2, \label{eq:full_psi_MSD_1}\\
    \tilde{\psi}_2(\tilde{\btheta}_2) &= \lambda \nu_3\psi_2\pa{\lambda^{-1}\tilde{\btheta}_2}+\frac{1}{2}\tilde{\eta}\pa{1-r_\eta\nu_3\lambda^{-1}}\tilde{\btheta}^2_2, \label{eq:full_psi_MSD_2}
\end{align}
where
\begin{equation}
    r_\eta = \frac{\eta}{\tilde{\eta}}, \qquad
    \nu_3 = \frac{\tilde{\eta}-1}{\eta-1},
    \label{eq:r_eta_and_nu_3}
\end{equation}
and with the transformed position $\tilde{\btheta}_2 = \lambda\btheta_2$ following from \autoref{eq:eff_lens_1}.

In the above equations, $\eta$ refers to the true physical scaling factor as defined in \autoref{eq:los_eta}, while $\tilde{\eta}$ refers to the quantity appearing with the lens model, which could be either fixed or left free, and which may differ from the true value of $\eta$. If we knew $\eta$ exactly, we could fix $\tilde{\eta}=\eta$, hence $r_\eta=\nu_3=1$. In reality, however, $\eta$ can never be known exactly. Even if the redshifts are measured extremely precisely, there will always be uncertainties from the cosmological model and matter on the line of sight, and this uncertainty will propagate into the profile of the second lens. 
%
\subsection{Mass-sheet degeneracy and time delays}
%
In \cite{Johnson_2026}, we showed that time-delay ratios are also preserved under a mismeasurement of $\eta$, and the time-delay function can be expressed as
\begin{align}
    \tilde{T}(\btheta, \tilde{\btheta}_2, \tilde{\btheta}_3)
    &= \lambda\bigg[\tilde{\tau}_{12} \phi_{12}(\btheta,\btheta_2) \nonumber \\
    & \qquad +\pa{\frac{\tilde{\eta}-1}{\eta-1}}\tilde{\tau}_{23}\phi_{23}(\btheta_2,\btheta_3)\bigg] \label{eq:time_delay_unsubstituted_MSD} \\
    &= \lambda\pa{\frac{\tilde{\tau}_{12}}{\tau_{12}}} \, T(\btheta,\btheta_2,\btheta_3).  \label{eq:tau12_factored_msd} 
\end{align}
$\tilde{T}(\btheta, \tilde{\btheta}_2, \tilde{\btheta}_3)$ refers to the time delay function which would be predicted from mass-sheet transformed lensing potentials $\tilde{\psi}_1(\btheta)$ and $\tilde{\psi}_2(\tilde{\btheta}_2)$.

Time delays therefore do not reduce the uncertainty in $\eta$. This mismeasurement will not bias the inferred Hubble constant, and the MSD-correction which is needed remains of the familiar form
\begin{equation}
    H_0 = \lambda^{-1}\pa{1-\kappa\h{LOS}_{12}}\tilde{H}_0. \label{eq:H0_msd}
\end{equation}
%
\section{Interpreting and exploiting mass-sheet constraints from double plane systems}
\label{sec:mass-sheet constraints}
%
\subsection{Flexible mass models}
\label{sec:flexible_mass_models}

Equations~\ref{eq:mst_1_k} and \ref{eq:mst_2_k} have the same basic structure -- a rescaling of the measured lens mass, and the addition of a constant (possibly negative) mass sheet. The rescaling is homogeneous, and can simply be absorbed by a redefinition of the Einstein radius of the lens. \Autoref{eq:mst_2_k} also rescales the argument of $\kappa_2$, but this rescaling can likewise be reabsorbed into the parameters of basic lens models, as is shown in \autoref{sec:power_law_msd_plane_2}. On the other hand, parametric lens mass profiles such as the elliptical power law \citep[EPL,][]{Tessore_2015} are qualitatively changed by the addition of a mass sheet, and thus their use artificially fixes the mass-sheet term, which must then be corrected for later in the analysis \citep{Schneider_2013}. A mass model which avoids such an artificial constraint must therefore be able to accommodate a mass-sheet-like term.

Now, while an infinite sheet is clearly unphysical, lensing observables are generally insensitive to the symmetric mass distribution at distances much larger than the Einstein radius. The addition of an extended (possibly negative) core component, approximately constant on this domain and then decreasing rapidly beyond $\theta\e{E}$, is essentially indistinguishable from a mass-sheet transformation, and a number of well-motivated mass models satisfy these criteria \citep[e.g.][]{Blum_2020,Blum_2021}. Adapting from \cite{Blum_2020}, we therefore consider mass profiles of the form
\begin{equation}
    \kappa\e{lens}(\btheta) = \kappa
\h{mod}(\btheta) + \zeta\kappa\e{c}(\btheta). 
    \label{eq:fid+core}
\end{equation}
$\kappa\e{mod}(\btheta)$ refers to some fiducial mass profile used in the modelling (often an EPL), while  $\zeta\kappa\e{c}(\btheta)$ refers to an additional mass component where $\zeta$ is a constant, and $\kappa\e{c}(\btheta) \approx 1$ on scales probed by the lens images, while dropping off at larger radii. We emphasise that the parameters of the $\kappa\e{mod}(\btheta)$ component of the true mass profile may differ from those inferred in the modelling, as the mass-sheet degeneracy in both planes will rescale this component. Nonetheless, sufficiently constraining lensing data will determine the functional form of the mass profile $\kappa\e{mod}(\btheta)$, leaving only the rescaling and the addition of $\zeta\kappa\e{c}(\btheta)$ unknown. Comparing to Eqs.~\ref{eq:mst_1_k} and~\ref{eq:mst_2_k}, we see that the profile in \autoref{eq:fid+core} is sufficiently general to fully accommodate the mass sheet transformations in both lens planes, while avoiding the unphysical behaviour of a truly infinite sheet. In the following, this will therefore be our working hypothesis, in other words, that \autoref{eq:fid+core}, where the functional form of $\kappa\e{mod}(\btheta)$ is identified from image constraints, is a sufficiently general description of the lenses we will consider.

\newpage

\subsection{Mass profile slopes}
\label{sec:cores_cusps_power_laws}

\begin{figure*}[t]
\centering
\includegraphics[width=\textwidth]{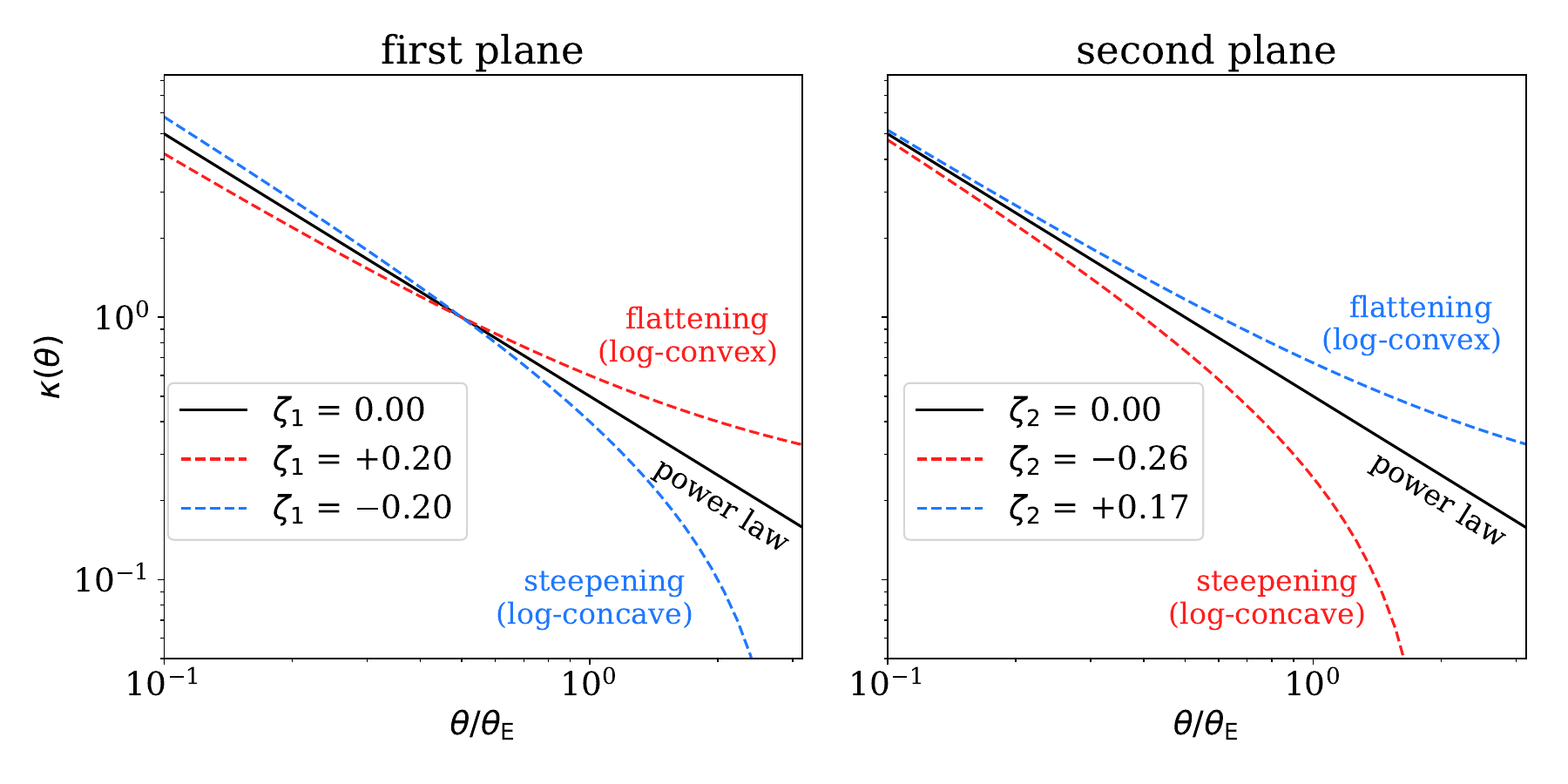}
\caption{Illustrative possibilities for the mass profiles in planes 1 and 2 of a double-plane lens, assuming $\eta = 1.021$ (our fiducial value for J1721), $\zeta_{12}=0$ and both planes have been modelled with a power law mass profile, following \autoref{eq:kappa1_msd}, \ref{eq:kappa2_msd} and \ref{eq:zeta_12_definition}. If, in reality, plane~1 features a positive mass-sheet component, $\zeta_1>0$, which would imply a log-convex profile which grows increasingly flat as $\theta$ increases relative to the modelled power law, the profile in plane 2 must be log-concave, growing increasingly steep as $\theta$ increases (the scenario in red). If instead plane 1 features a negative mass-sheet component, this relationship is reversed (the scenario in blue). \vspace{1em}}
\label{fig:core_and_cusp}
\end{figure*}

The construction of the expressions in Eqs.~\ref{eq:mst_1_k}, \ref{eq:mst_2_k} and \ref{eq:fid+core} are suggestive of a distinct sheet-like component added to a conventional mass profile. In reality, the simultaneous rescaling and addition of the sheet acts as a global change to the density slope of the mass model.  
At larger radii, the additional $(1-\lambda)$ term dominates, while closer to the centre of the lens, it is the effect of the $\lambda$ which multiplies the base profile which dominates. The overall effect is that $\lambda>1$ leads to profiles which are increasingly steep at higher $\theta$, while $\lambda<1$ gives a profile which flattens as $\theta$ increases.\footnote{Note that, this is only true for the region when $\kappa\e{c}\approx1$. For larger radii and finite sheet-like components, $\lambda$ will be suppressed, as is very nicely illustrated in Figure 1 of \cite{Blum_2020}.} A mass-sheet transform is therefore best understood as a change to the \textit{derivative} of the logarithmic mass slope.

If the mass-sheet component $\zeta$ in \autoref{eq:fid+core} is artificially fixed to zero, and a power law is used as the fiducial profile, then a true $\zeta>0$ would imply a true profile which was log-convex with respect to the power law model, and a true $\zeta<0$ would imply a true profile which was log-concave. This is explored further in \autoref{sec:what_we_learn_image_positions}, and illustrated in \autoref{fig:core_and_cusp}.

In the following, we will describe profiles as ``steepening'' or ``flattening'',\footnote{Steepening and flattening profiles are sometimes referred to as ``cuspy'' and ``cored'', respectively, in the lensing literature. We preferred to avoid these denominations because the value of $\lambda$ does not actually change the cuspy or cored nature of a density profile -- a power law always dominates a mass sheet close to the centre.} and this should be understood with respect to the fiducial model used in the modelling, in the region near the image positions. Importantly, quantities such as $\lambda$ in \autoref{eq:mst_1_k} or $\zeta$ in \autoref{eq:fid+core} are also defined with respect to this fiducial mass model. Constraints on $\lambda$ or $\zeta$ found with an EPL mass model cannot therefore be robustly compared to those found with e.g. a composite model. While the message remains the same regardless of this choice, for consistency we will assume that our base model is an EPL wherever necessary, with a convergence given by
\begin{equation}
    \kappa\e{EPL}(\vartheta) = \frac{3-\gamma}{2}\left(\frac{\theta\e{E}}{\vartheta}\right)^{\gamma-1},
    \label{eq:EPL_convergence}
\end{equation}
where $\gamma$ is the power law slope, $\theta\e{E}$ is the Einstein radius, i.e. the characteristic radius corresponding to the typical separation between the centre of the lens and the images which form around it, and $\vartheta=\theta\sqrt{q\cos^2\varphi+(1/q)\sin^2\varphi}$ is the elliptical radius (equal to $\theta$ for a circularly-symmetric power law), $q$ is the axis ratio, $\varphi$ is measured relative to the major axis, and $\theta = |\btheta|$.

\subsection{What we learn from image positions}
\label{sec:what_we_learn_image_positions}

The lesson in \cite{Schneider_2014} is that, if $\eta$ is bounded, while $\zeta$ is not observable in either lens plane individually, its presence in one lens cannot be mass-sheet transformed away without affecting what is inferred about the mass in the other lens. To make this explicit, consider the mass-sheet transformed versions of \autoref{eq:fid+core}. Substituting \autoref{eq:fid+core} into Eqs.~\ref{eq:mst_1_k} and \ref{eq:mst_2_k}, omitting the $\kappa\e{c}(\btheta)$ factors which by definition have no effect on our domain of interest, we obtain
\begin{align}
    \tilde{\kappa}\h{lens}_1(\btheta) &= \lambda\kappa\h{mod}_1(\btheta) + \lambda\zeta_1+\pa{1-\lambda}  \label{eq:kappa1_msd}\\ 
    &= \tilde{\kappa}\h{mod}_1(\btheta)+\tilde{\zeta}_1, \label{eq:mst_1_k_core} 
\end{align}
and
\begin{align}
    \tilde{\kappa}\h{lens}_2(\tilde{\btheta}_2)  &= \nu_3\lambda^{-1}\kappa_2\pa{\lambda^{-1}\tilde{\btheta}_2} \nonumber \\ 
    &\qquad + \nu_3\lambda^{-1}\zeta_2 +\tilde{\eta}\pa{1-r_\eta\nu_3\lambda^{-1}} \label{eq:kappa2_msd} \\
    &= \tilde{\kappa}\h{mod}_2(\btheta)+\tilde{\zeta}_2. \label{eq:mst_2_k_core}
\end{align}
The double-plane MSD ensures that we have the freedom to model our systems with Eqs.~\ref{eq:mst_1_k_core} and \ref{eq:mst_2_k_core}, and set $\lambda$ and $\tilde{\eta}$ to whatever we would like without changing any imaging observables. As such, we may choose to omit $\tilde{\zeta}_1$ and $\tilde{\zeta}_2$ from the model entirely, modelling both lenses as EPLs (or any other profile of our choice). If we make this choice in the first plane, we have implicitly fixed the relationships between $\lambda$ and $\zeta_1$,
\begin{gather}
    \tilde{\zeta}_1 = 0 \implies \zeta_1 = 1 - \lambda^{-1}. \label{eq:implied_lambda}
\end{gather}
Making this choice in the second plane, 
\begin{gather}
    \tilde{\zeta}_2 = 0 \implies  \nu_3\lambda^{-1}\zeta_2+\tilde{\eta}\pa{1-r_\eta\nu_3\lambda^{-1}} =0. \label{eq:zeta_2=0}
\end{gather}
The value of $\lambda$ is fixed by \autoref{eq:implied_lambda}, and since $\eta$ is a physical quantity, the value of $\tilde{\eta}$ is fixed by the choice to set $\tilde{\zeta}_2=0$ and model plane 2 with a power law model. We can solve \autoref{eq:zeta_2=0} to write
\begin{gather}
    \zeta_{12}\equiv\eta\zeta_1+\zeta_2-\zeta_1\zeta_2 = \frac{\tilde{\eta}-\eta}{\tilde{\eta}-1}. \label{eq:zeta_12_definition}
\end{gather}
Here, the qualitative differences between single and double-plane lensing become apparent. From single plane lensing without time delays, the mass-sheet component $\zeta$ can never be constrained. Even when time delays are present, only a prior on $H_0$ or independent probes can break this degeneracy. In double-plane lensing, on the other hand, even without time delays, the quantity $\zeta_{12}$ as defined in \autoref{eq:zeta_12_definition} can be inferred by comparing the measured $\tilde{\eta}$ value and a reference $\eta$ value from other cosmological surveys.

\begin{figure*}
    \centering
    \includegraphics[width=1\textwidth]{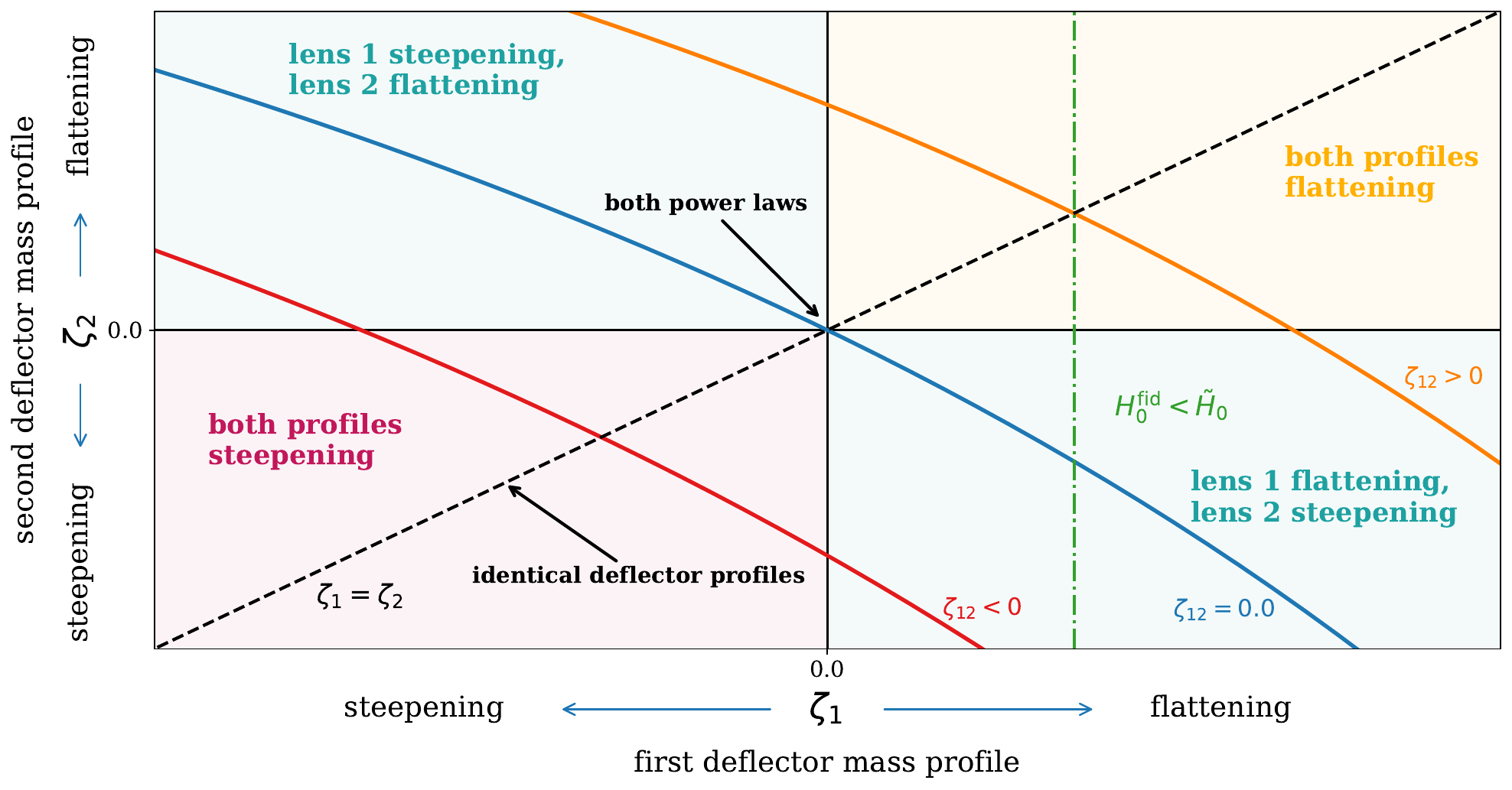}
    \caption{What we learn from observations of a double plane lens. Imaging constraints and a reference $\eta$ value (i.e. the relative expansion history) provide a measurement of $\zeta_{12}$, fixing us to a line of degeneracy (as shown by the curved solid coloured lines), and thus limits the combination of lens profile shapes (with respect to the fiducial profile). While we do not fix the profile of either lens, we can restrict the combinations of the profiles of the two lenses. Requiring that $\zeta_1=\zeta_2$ (shown as the black dotted line), or measuring time delays and comparing to a value of $H_0$ (shown as the green dash-dotted line), would each fix the values of both $\zeta_1$ and $\zeta_2$. Any disagreement between these constraints implies a flaw in one of their underlying assumptions. 
    }
    \label{fig:theory}
\end{figure*}

The value of such a measurement is illustrated by \autoref{fig:theory}. The solid curved lines correspond to different measurements of $\zeta_{12}$. From imaging constraints alone, we eliminate large portions of the $(\zeta_1,\zeta_2)$ parameter space - for example, a positive value of $\zeta_{12}$ is incompatible with both lenses being steepening compared to a power law, and if $\zeta_{12}=0$, the lenses must both be power laws, or have opposite-sign mass-sheet components. If the profiles are restricted to be similar, i.e. as having a limited dispersion in $\zeta_i$, then we fix ourselves to lie near the black dotted line, and can thus use our constraint on $\zeta_{12}$ to constrain the profile of both lenses. 

\subsection{What we learn from time delays}
\label{sec:what_we_learn_time_delays}

\begin{figure*}
    \centering
    \includegraphics[width=0.97\textwidth]{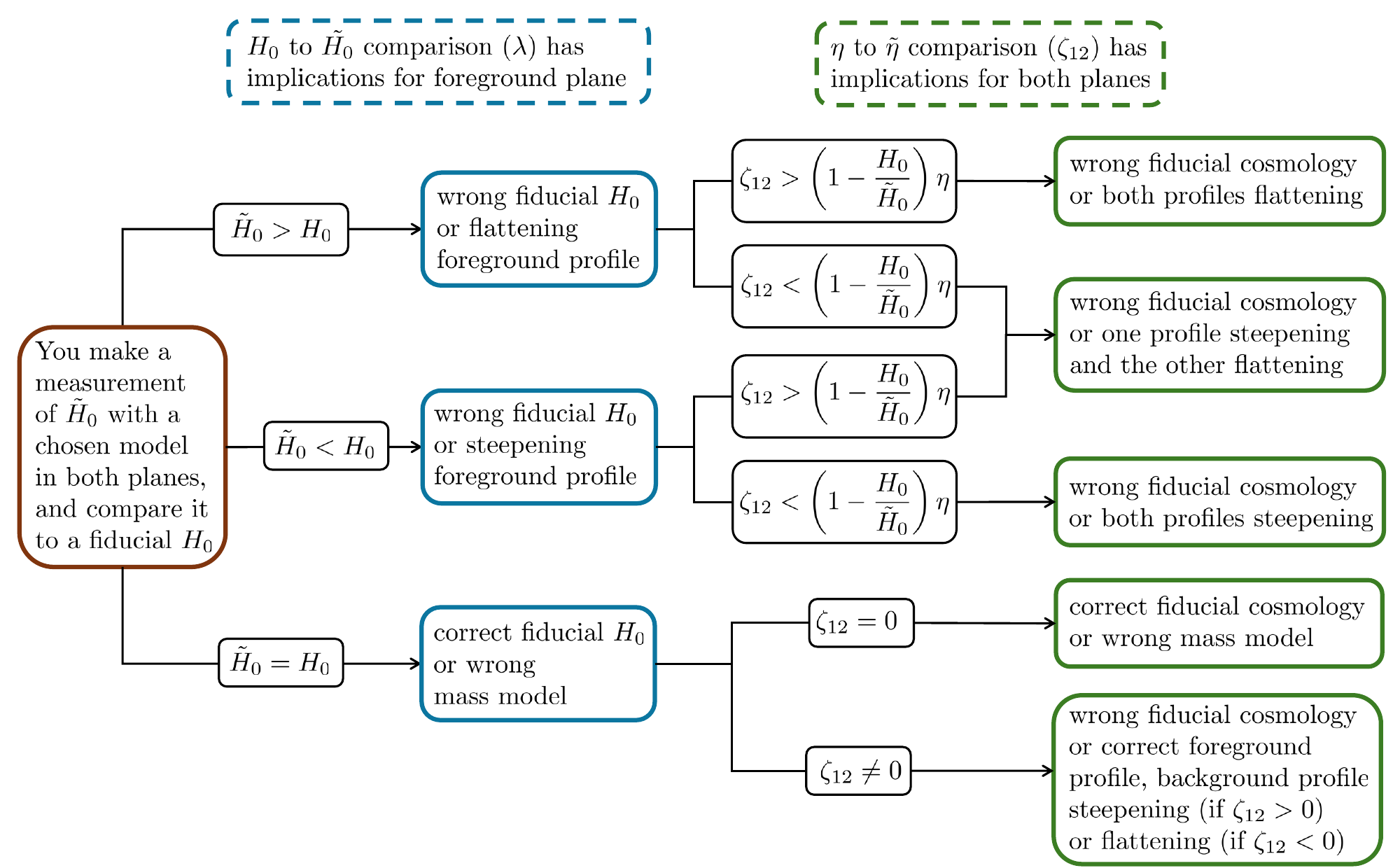}
    \caption{What can be learnt from a double-plane system with time delays. A $\tilde{H}_0$ measurement is sensitive to $H_0$ and the mass profile in the foreground plane, while a $\tilde{\eta}$ measurement is sensitive to the relative expansion history and the mass profiles in both planes. If you trust a certain fiducial cosmology, this has implications for the mass profiles of the two lens galaxies. 
    }
    \label{fig:tree_diagram}
\end{figure*}

Suppose further that relative time delays are measured for such a system, in either plane. We saw in \autoref{sec:double_plane_mst} that the combination of these time delays and mass modelling constrains the combination $\tilde{H}_0=\lambda H_0$. In turn, the choice to set $\tilde{\zeta}_1=0$ means that $\lambda$ will relate to the true value of $\zeta_1$ via \autoref{eq:implied_lambda}, i.e.
\begin{equation}
    \zeta_1 = 1 - \frac{H_0}{\tilde{H}_0}.
\end{equation}
Assuming a fiducial value $H\h{fid}_0$ \citep[from][for example]{Planck_2020} therefore fixes us to a single $\lambda$ (and hence $\zeta_1$) value, and in turn a value of $\zeta_2$ via $\zeta_{12}$,
\begin{equation}
    \zeta_2 = \lambda\zeta_{12}-\lambda\eta+\eta. 
\end{equation}
It is straightforward to see from \autoref{fig:theory} that the measurement of $\zeta_{12}$ via a reference $\eta$ value, the requirement that $\zeta_1 \approx \zeta_2$, and the assumption of a certain value of $H\h{fid}_0$, need not agree. The possibilities are summarised in \autoref{fig:tree_diagram}.

Real measurements are, of course, not exact, and various sources of uncertainty (such as from the measurement of $\zeta_{12}$, the line of sight convergence, the value of $H\h{fid}_0$, the measured time delays) will each contribute to blurring the lines between these possibilities. Nonetheless, the method grants access to novel constraints which are not possible from the images and time delays of single plane lenses, regardless of their precision and abundance.
%
\subsection{What we learn from stellar kinematics}
\label{sec:kinematics}
%
As mentioned before, the conventional method of constraining the internal mass-sheet component of gravitational lenses relies on measurements of the velocity dispersion of the stars in the lens galaxy, which depends of the mass of that galaxy. By comparing the model-predicted dispersion to the observation, we measure the difference in normalisation, and hence disentangle the lens mass from a combination of the internal and external mass-sheet component \citep{Birrer_2020,Birrer_2022}. 

These techniques have developed extensively over the last decade, accounting for the nuance of inclination and anisotropy, but to leading order, what such a measurement constrains is the rescaling of the displacement angle, up to a correction for finite core effects, which we will neglect in this analysis \citep[see e.g.][]{Yildirim_2021,Shajib_2023}. A key result in \cite{teodori_2026} is that this result remains true even for the subsequent lens planes, in which the velocity dispersion is measured from a reconstruction of the lens galaxy according to the foreground lens model/s. The constraints in plane 2 (and for subsequent lens planes when they are present) are sensitive not only to any mass-sheet in that plane, but additionally to any in the foreground plane/s which would affect the predicted mass in the second plane. 

Adapting the result from \cite{teodori_2026} to our context, what is constrained from velocity dispersion measurements of the two lens galaxies is
\begin{align}
    V_1 &\equiv \frac{\tilde{\sigma}^2_1}{\bar{\sigma}^2_1} = \lambda\frac{1-\kappa_{12}}{1-\kappa_2}, \label{eq:v1} \\
    V_2 &\equiv \frac{\tilde{\sigma}^2_2}{\bar{\sigma}^2_2} = \frac{\tilde{\eta}-1}{\eta-1}\frac{1-\kappa_{23}}{1-\kappa_3}, \label{eq:v2}
\end{align}
where $\tilde{\sigma}^2_l$ is the velocity dispersion in plane $l$, predicted from the lens model, and $\bar{\sigma}^2_l$ is the actual velocity dispersion of the lens galaxy, which would be measured. Using the expressions from Eqs.~\ref{eq:implied_lambda} and \ref{eq:zeta_12_definition}, with a bit of algebra, we could write 
\begin{align}
    V_1 & = \frac{1-\kappa_{12}}{1-\kappa_2}\frac{1}{1-\zeta_1}, \\
    V_2 &= \frac{1-\kappa_{23}}{1-\kappa_3}\frac{\pa{1-\zeta_1}+\zeta_1\tilde{\eta}}{(1-\zeta_1)(1-\zeta_2)}.
\end{align}
Note that, when the $\zeta$s are close to 0, $V_2 \approx \zeta_{12}+1$, meaning that a measurement of $V_2$ and a comparison between $\tilde{\eta}$ and $\eta$ give similar information in this regime.

\subsection{Complications from the line of sight}
\label{sec:los_complications}

The distinction between matter on the line of sight and that in the main lens is on the one hand artificial, and on the other hand important. The weak lensing corrections appearing in Eqs.~\ref{eq:multi-plane_lens}, \ref{eq:alpha1_los}, \ref{eq:alpha2_los} and elsewhere are naturally absorbed into the corresponding lens profiles (or equivalently into the associated angular diameter distances). One option is therefore to simply understand $\zeta_1$ and $\zeta_2$ as they appear in $\zeta_{12}$, as combinations of line-of-sight convergences and intrinsic sheet-like components $\bar{\zeta}_1$ and $\bar{\zeta}_2$ of the two lens galaxies, 
\begin{align}
    \zeta_1 &= \frac{(1-\kappa_{12})(1-\kappa_1)}{1-\kappa_2}\bar{\zeta}_1, \label{eq:zeta1_with_los} \\
    \zeta_2 &= \frac{(1-\kappa_{23})(1-\kappa_2)}{1-\kappa_3}\bar{\zeta}_2, \label{eq:zeta2_with_los}
\end{align}
and to restrict ourselves to comments about these combinations. While a very natural choice when modelling a given double-plane system, this limits the conclusions we can draw from our $\zeta_{12}$ measurement about the mass distribution of the two galaxies themselves, as well as our ability to compare constraints from multiple systems.

The line of sight also complicates our constraint of $\zeta_{12}$ itself. Even if the background cosmology was known perfectly, the parameter $\eta$ appearing in \autoref{eq:zeta_12_definition} also contains a contribution from line-of-sight convergence terms, \autoref{eq:los_eta}, which expected to be small but not negligible~\citep{Johnson_2025}. 

To account for these contributions, each convergence term and its uncertainty should be estimated, which can be done by comparing the lens environment to simulated lines of sight within large weak lensing simulations \citep[see e.g.][]{Suyu_2010,Rusu_2017,Wells_2023,johnson_2026b}.

\section{Application to J1721+8842}
\label{sec:zigzag_application}

\begin{figure}
    \centering
    \includegraphics[width=\columnwidth]{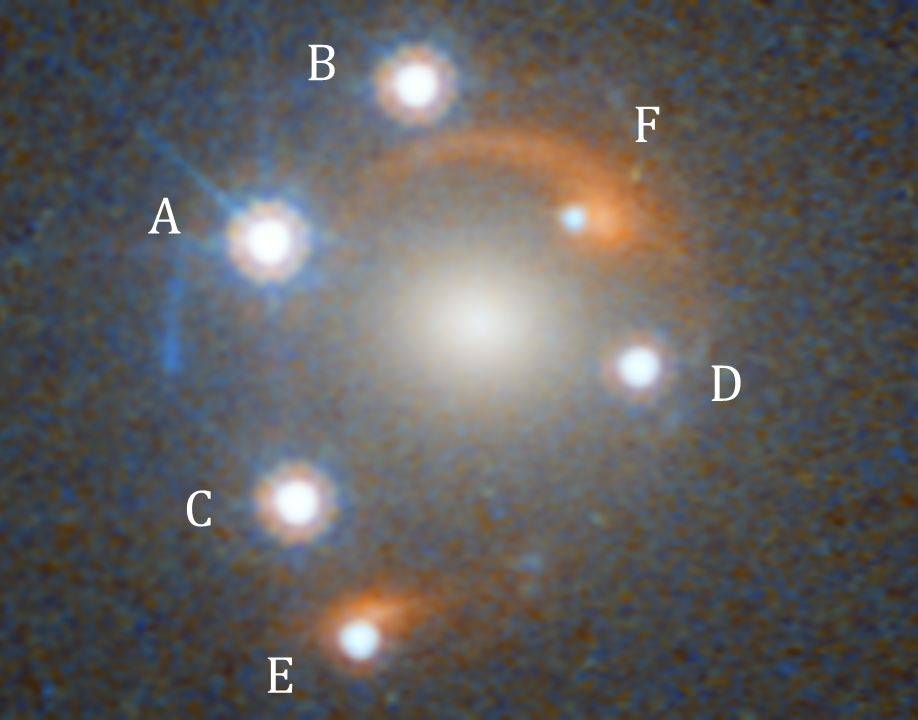}
    \caption{HST/WFC3 composite image of J1721, the Einstein zigzag lens, produced with the help of {\sc STCI},\textsuperscript{5} showing the 6 image positions from which our constraints are derived. \vspace{1em}}
    \label{fig:zigzag}
\end{figure}

The ``Einstein zig-zag lens'', J1721+8842, shown in \autoref{fig:zigzag}, is a spectacular compound lens, featuring a sextuply imaged quasar at $z=2.382$ whose light is deflected by galaxies at redshifts $z_1=0.184$ and $z_2=1.885$, with light from the intermediate galaxy at $z_2$ also lensed to multiple extended arcs by the first lens~\citep{Lemon_2018, Dux_2025}. The variability of the quasar and position of the system in the sky have enabled time-delay measurements between all 6 images, and the exceedingly rare image configuration is highly sensitive to the mass profiles of both deflectors \citep{Collett_2016}. 

In the following, we describe the observation and modelling of this system, used to obtain a measurement of $\tilde{\eta}$, and therefore $\zeta_{12}$. In many ways, this system is an ideal candidate for a first inference of $\zeta_{12}$, with its abundant imaging information, and its potential use in $H_0$ measurements, allowing comparisons of the type described in \autoref{sec:what_we_learn_image_positions}. It is also a rare example of a DPL in which both planes strongly lens the source light, and is therefore more sensitive to the background lens' mass distribution. With upcoming velocity dispersion measurements of the second deflector plane and high-resolution JWST imaging, we anticipate that these constraints will be further improved.
\footnotetext{Space-telescope Color Optimization and Processing Engine, \url{https://github.com/astroskylee/STCI}.}

\subsection{Lensing observables}

In this work, we use the positions of the six quasar images as lensing observables. The HST images of this system contain significantly more lensing information, e.g. in the extended arcs at redshift \ztwo and \zs, but the unique lens configuration means that the point source images alone suffice for the exploratory, ``proof-of-concept'' goals of this work. For computational efficiency, we refrain from modelling the extended sources, which we leave for an upcoming work.

Astrometric measurements are obtained with the \textsc{STARRED} two-channel forward-modelling code \citep{Michalewicz_2023, Millon_2024}, applied to the \textit{HST}/F814W and F475W observations (GO-15652). For each band, the PSF is reconstructed empirically by jointly fitting the five brightest quasar images. We also tested PSF models derived from field stars, but these led to less accurate reconstructions of the quasar PSF, likely due to field distortions and color dependence of the PSF across the filter bandpass. \textsc{STARRED} is then used to obtain a high-resolution model of the system by jointly fitting the positions and amplitudes of the point-like components together with the extended emission (lensed arcs, foreground galaxy, etc.). The resulting astrometry for each band is presented in \autoref{tab:astrometry}, alongside the corresponding measurements from \textit{Gaia} \citep{Gaia_2018}. We find excellent agreement with \textit{Gaia}, with an RMS difference below 0.5\,mas for the five detected images. When modelling with these positions, we increase the astrometric uncertainty to 2\,mas, to account for the additional uncertainty from millilensing by substructures. We return to this point in \autoref{sec:caveats}.

\begin{table*}
\centering
\begin{tabular}{lcccccc}
\toprule
 & A & B & C & D & E & F \\
\midrule
HST RA [arcsec] & -0.4297 ± 0.0002 & 1.5023 ± 0.0003 & -2.0028 ± 0.0001 & 0.9674 ± 0.0002 & -2.5349 ± 0.0002 & 1.5733 ± 0.0005 \\
HST Dec [arcsec] & 1.7117 ± 0.0011 & 1.6739 ± 0.0002 & -0.0489 ± 0.0004 & -1.5101 ± 0.0003 & -1.2920 ± 0.0005 & -0.1824 ± 0.0003 \\
Gaia RA [arcsec] & -0.4301 & 1.5028 & -2.0026 & 0.9680 & -2.5357 &  -  \\
Gaia Dec [arcsec] & 1.7109 & 1.6740 & -0.0492 & -1.5102 & -1.2920 &  -  \\
\bottomrule
\end{tabular}
\caption{Comparison of measured mean F814W and F475W HST and Gaia astrometry for the point source images of J1721 (as labelled in \autoref{fig:zigzag}).}
\label{tab:astrometry}
\end{table*} 

\subsection{Lens model}
\label{subsec:lensmodel_J1721}

We model the foreground lens galaxy with an EPL+shear at \zone, with an SIS for the background lens galaxy at \ztwo. The 4 perturbing galaxies identified in \cite{Schmidt2025} are modelled with SIS profiles, whose parameters are fixed to the best-estimate values inferred in \cite{Schmidt2025} (see \autoref{sec:caveats}). We leave $\tilde{\eta}$ free to vary, and so the mass model contains 12 free parameters in total.

For each image observed position $\btheta^{(i)}_{1}$, we compute the corresponding source position $\btheta^{(i)}_{3}$ in the \zs source plane from ray tracing according to the lens model parameters $\bm{\xi}_{\rm lens}$. These computations are carried out with the lens modelling code {\sc lenstronomy} \citep{Birrer_2021b}. The parameter posterior distributions are obtained from Bayes theorem,
\begin{equation}
    P(\bm{\xi}\e{lens}, \btheta_3, \Theta| \btheta_1) = \mathcal{L}( \btheta_1 | \bm{\xi}\e{lens}, \btheta_3, \Theta) \mathcal{P}(\bm{\xi}\e{lens}, \btheta_3, \Theta),
\end{equation}
where $\bar{\btheta}_{3}$ is the mean source position produced by the model, $\sigma_{\theta_3}$ is the astrometric uncertainties, $\mathcal{L}$ is the likelihood and $\mathcal{P}(\bm \xi_{\rm lens}, \bm{\theta_3}, \Theta)$ is a non-informative uniform prior on all lens
and source position parameters. 
%
\subsection{Estimating the line of sight contribution}
\label{sec:zigzag_los_estimate}
\begin{figure}
    \centering
    \includegraphics[width=1\columnwidth]{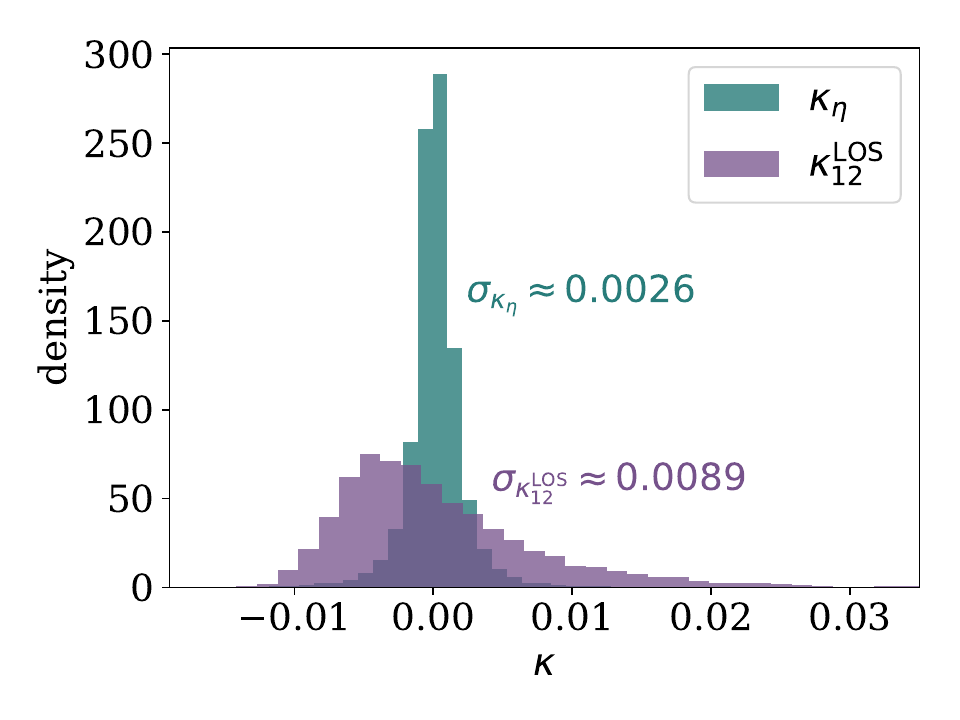}
    \caption{Estimated line-of-sight contribution to the uncertainty in $\eta$, $\kappa_\eta$, and for a hypothetical $H_0$ measurement, $\kappa\h{LOS}_{12}$, for J1721, using the methods described in \autoref{sec:zigzag_los_estimate}.\vspace{1em}}
    \label{fig:1721_los}
\end{figure}
Properly estimating the contribution of the line-of-sight terms in Eqs.~\ref{eq:los_eta}, \ref{eq:zeta1_with_los}, and \ref{eq:zeta2_with_los} will require wide-field observations and weighted number count techniques, as applied in \cite{johnson_2026b}. For the purposes of this work however, we restrict ourselves to ``naive'' estimates, drawing simulated lines of sight to obtain a posterior which is not weighted by observations of the lens environment itself. This posterior will capture the uncertainty coming from the line-of-sight convergence, but not a systematic shift due to any over- or under-density we may observe in galaxy counts.

To this end, we make use of the RayGalGroupSims~\citep{Breton:2018wzk},
a dark-matter simulation in which fully relativistic ray-tracing has been carried out, producing publicly-available HEALPix maps of convergence and shear for a range of redshifts\footnote{\url{https://cosmo.obspm.fr/public-datasets/raygalgroupsims-relativistic-halo-catalogs}}. The simulation is based on the $N$-body code RAMSES~\citep{Teyssier_2002, Guillet_2011} using the \textsc{Magrathea} library~\citep{reverdy2014propagation, Breton, Breton_2022}.

Following the procedure in \cite{Johnson_2025}, we extract lines of sight $\uvect{n}$ at random from the simulation, and interpolate between convergence values to obtain $\kappa(z_i,\uvect{n})$ at arbitrary redshift $z_i$, from which we can calculate $\kappa_{ij}$ for $i,j\in\{0,1,2,3\}$. We then use \autoref{eq:kappa_LOS} to estimate the distributions of $\kappa\h{LOS}_{13}$ and $\kappa\h{LOS}_{23}$ for the redshifts associated with J1721+8842.  The resulting distributions of $\kappa_{\eta}$ and $\kappa\h{LOS}_{12}$ for J1721 are shown in \autoref{fig:1721_los}.
\subsection{Constraining the population dispersion with single-plane time-delay lenses}
\label{sec:pop_dispersion}
As discussed in \autoref{sec:MSD}, measurements of $H_0$ from time delay cosmography without a correction for the mass-sheet degeneracy are in fact measurements of $\pa[2]{1~-~\zeta_1}^{-1}\pa[2]{1~-~\kappa\h{LOS}_{12}}^{-1}  H_0$, where the value of $\zeta_1$ is specific to that lens and the mass profile with which it is modelled, and $\kappa\h{LOS}_{12}$ is specific to that line of sight. 

Suppose we have a sample of time-delay lenses, and from this, a collection of measurements of $\pa[2]{1~-~\zeta_1}^{-1}\pa[2]{1~-~\kappa\h{LOS}_{12}}^{-1}~H_0$ for each lens system. Suppose we also have estimates for each $\kappa\h{LOS}_{12}$. As the value of $H_0$ doesn't change, any scatter in these values beyond what is explained by the measurement uncertainty and $\kappa\h{LOS}_{12}$ must come from an intrinsic scatter in $\zeta$. Thus, if we assume that our sample of lenses is drawn from some common underlying distribution, we can use this sample to constrain the dispersion of that population.

To obtain such a constraint, we make use of the sample of lenses in \cite{Millon_2020}. Each of these lenses were modelled using an elliptical power law, matching our choice for J1721, and already incorporate a correction for the uncertainty from the line of sight. Under various simplifying choices, assuming that the underlying population of $\zeta$ values follow a Gaussian distribution with standard deviation $\sigma_{\bar{\zeta}}$, we can place a $1\sigma$ upper bound on $\sigma_{\bar{\zeta}}$ of 0.039. The details of this analysis and the necessary assumptions and simplifications can be found in \autoref{sec:dispersion_estimate}. 
\subsection{Results, comparison to the fiducial cosmology, and implications for the mass-sheet degeneracy}
\label{sec:zigzag_comparison_to_prior}
\begin{figure}
    \centering
    \includegraphics[width=1\columnwidth]{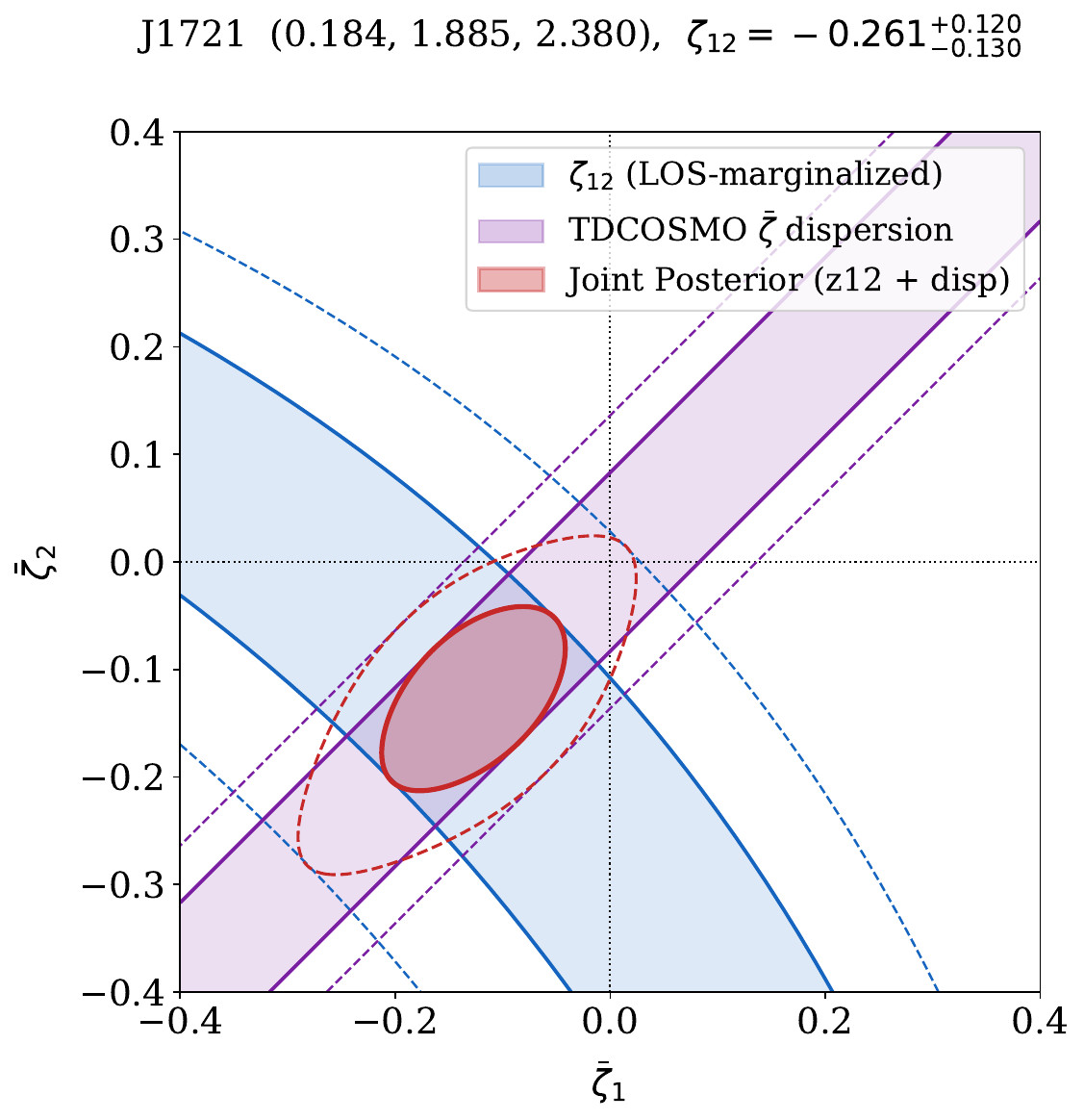}
    \caption{Constraints on the mass-sheet components of J1721. In blue, we show the constraints from $\zeta_{12}$, which is a comparison between our $\tilde{\eta}$ measurement, a fiducial $\Omega\e{m}$ value from \cite{Planck_2020}, and an error budget for the line of sight. In purple, we show the the constraints on the mass-sheet population scatter from \autoref{sec:pop_dispersion} for the lenses in \cite{Millon_2020}, using and the assumption that the lens planes in J1721 are drawn from this same population. We show the joint constraints in red. \vspace{1em}}
    \label{fig:zigzag_constraints}
\end{figure}
From the aforementioned data and model, we measure $\tilde{\eta} = 1.0166 ^{+0.0011}_{-0.0011}$.
As discussed in \autoref{sec:what_we_learn_image_positions}, a measurement of $\tilde{\eta}$ alone is not enough to constrain a value of $\zeta_{12}$. To do so, we must place a prior on $\eta=(1-\kappa_\eta)\bar{\eta}$, via a prior on both the cosmology and on $\kappa_\eta$. For $\kappa_\eta$, we use the estimate of $\sigma_{\kappa_{\eta}} = 0.0026$ from \autoref{sec:zigzag_los_estimate}. For our prior on $\bar{\eta}$, we use $\Omega\e{m}=0.315\pm0.007$ for a Flat $\Lambda$CDM universe from \cite{Planck_2020}, which together give us $\eta = 1.0210 ^{+0.0013}_{-0.0015}$. Combining this with the measured $\tilde{\eta}$, we obtain $\zeta_{12} = -0.22 ^{+0.12}_{-0.13}$.

The parameter space of $\bar{\zeta}_1$ and $\bar{\zeta}_2$, the intrinsic mass-sheet components of the two lenses, is shown in \autoref{fig:zigzag_constraints}. In blue, we show the constraints implied by the measurement of $\zeta_{12}$. This result alone disfavours both lenses being power laws. If we incorporate the profile dispersion constraint from \autoref{sec:pop_dispersion}, shown in purple, the two constraints imply $\bar{\zeta}_1=-0.129^{+0.061}_{-0.063}$ and $\bar{\zeta}_2=-0.0129^{+0.061}_{-0.064}$, i.e. two log-concave profiles.
\subsection{Caveats: lens model flexibility, population scatter, and the astrometric uncertainty from millilensing}
\label{sec:caveats}

The results in \autoref{sec:zigzag_comparison_to_prior} are a compelling proof of concept -- with only the quasar image positions and a fiducial $\Omega\e{m}$ value, we can restrict the MSD parameter space of the system, without any need for velocity dispersion measurements, an $H_0$ prior, or restrictive assumptions about possible mass-sheet components in either plane. Nonetheless, these results come with some important caveats. The first is that, by fixing the parameters of the SIS perturber galaxies, the parameter space of the model is not fully sampled, which may lead us to underestimate the $\tilde{\eta}$ uncertainty. With only point source images, however, we do not have enough constraints to leave the perturber parameters to vary. We nevertheless note that these extra degrees of freedom (e.g. power-law slope and ellipticity of the second perturber) can be constrained from the modelling of the high-resolution images, as demonstrated in \citet{Schmidt2025}.

While $\zeta_{12}$ alone has implications for the MSD, we cannot say anything about $\bar{\zeta}_1$ and $\bar{\zeta}_2$ individually without an additional constraint, which in \autoref{fig:zigzag_constraints} came from the assumption that we could place a limit on the scatter in $\bar{\zeta}$ from the sample in \cite{Millon_2020}. We will return to this point in \autoref{sec:discussion}, but it is almost certainly too strong an assumption to be fully robust. Additionally, the scatter we find is somewhat smaller than that found in e.g. \cite{TDCOSMO_2026} for a larger sample of lenses when velocity dispersion data is incorporated.
\subsubsection{The astrometric uncertainty}
Finally, our uncertainty budget for the millilensing by line-of-sight subhalos, although a reasonable estimate for typical time-delay lenses, is almost certainly an underestimate for this particular system, in which the point source positions are highly sensitive to changes in the mass model. To test this, having obtained the parameters of the macromodel in \autoref{subsec:lensmodel_J1721}, we proceed according to the method from \cite{Gilman_2020}. The program \texttt{pyHalo} \citep{Gilman_2020}\footnote{\url{https://github.com/dangilman/pyHalo}} is used to simulate a population of CDM subhalos along the line of sight, for which we assume a standard $\Lambda$CDM cosmology \citep{Planck_2020}, and subhalo parameters, in particular a $\sigma_{\rm{sub}}=0.1 \frac{1}{\rm{kpc}^{-2}}$ amplitude of the subhalo mass function, as expected from CDM predictions, and a Sheth-Tormen mass function for the LOS halos \citep{Sheth_2001}. 


The halos are sampled on a cone opening of 20 arcseconds, centred around the macro-model light-path. The additional mass introduced by these halos, which would otherwise bias the total convergence, is subtracted from the mass model as a negative mass sheet correction. Once the population is simulated, the image positions are ray-traced backwards to the source plane, and their average position defines the millilensing-perturbed source position. This is repeated over many 
realisations of the subhalo population, from which a distribution of source positions is obtained. The average distances between these positions and the source position from the smooth macro-model serves as the estimated astrometric uncertainty from millilensing, which we find to be $\sigma_{\theta_3} = 9.1$mas. 
As stated above, this estimate is higher than that for most systems, and comes from the highly sensitive 6-image configuration of J1721. With this value, however, $\tilde{\eta}$ is effectively unconstrained. We therefore retain the more typical 2mas estimate, but emphasise once again that our results should therefore be taken only as a proof of concept, and more robust constraints will follow from the incorporation of the extended source images.
%
\subsection{What will stellar kinematics bring?}
\begin{figure}
    \centering
    \includegraphics[width=1\columnwidth]{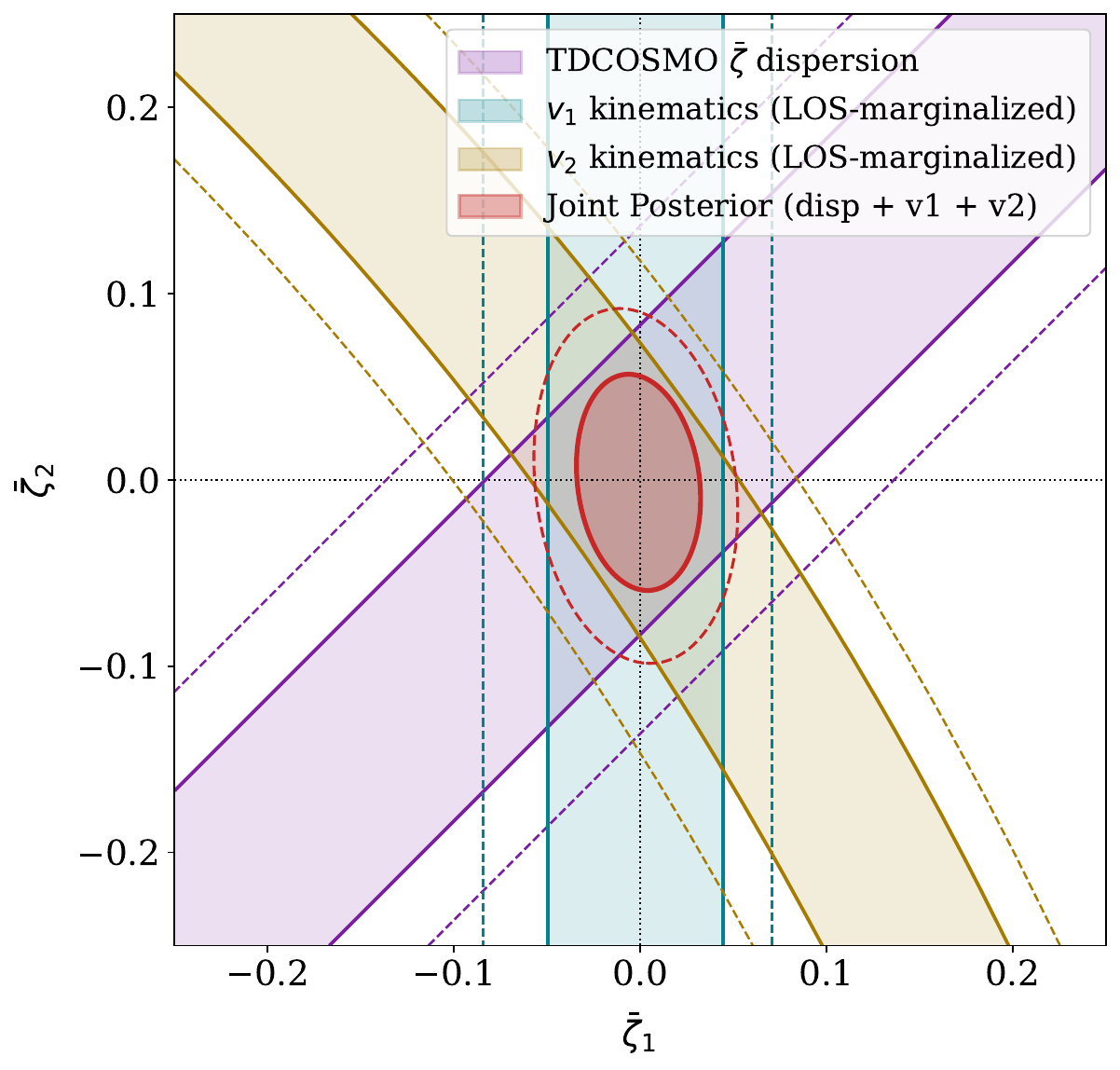}
    \caption{Hypothetical constraints for J1721 with a 3\% velocity dispersion measurement in plane 1 and a 5\% measurement in plane 2, with our measurement of $\tilde{\eta} = 1.0166 ^{+0.0011}_{-0.0011}$ and our population dispersion constraint from \autoref{sec:pop_dispersion} shown in purple, and the joint constraints in red. \vspace{1em}}
    \label{fig:hypothetical_zigzag_constraints_with_kinematics}
\end{figure}
The measurement of $\tilde{\eta}$, as well as enabling a $\zeta_{12}$ constraint, will also be crucial for interpreting upcoming JWST/Nirspec resolved stellar kinematics measurements of the second plane from JWST GO programme 10159 (PI: M. Millon)\footnote{\url{https://www.stsci.edu/jwst/science-execution/program-information?id=10159}}. In \autoref{fig:hypothetical_zigzag_constraints_with_kinematics}, we show the constraints possible in the $(\zeta_1,\zeta_2)$ space when we combine our population constraint from \autoref{sec:pop_dispersion}, our estimated line-of-sight contribution from \autoref{sec:zigzag_los_estimate}, and hypothetical measurements of $V_1$ (shown in teal) and $V_2$ (shown in ochre), following \autoref{eq:v1}, \ref{eq:v2}. For the sake of the forecast, we assume both $V_1$ and $V_2$ are equal to 1.0 (i.e. that both lenses are in fact power laws), with uncertainties of 3\% and 5\% respectively. The combination of $V_1$ and $V_2$ would allow us to constrain $\zeta_1$ and $\zeta_2$ with 1$\sigma$ uncertainties of 0.03 and 0.07 respectively, improving to 0.02 and 0.04 when the population prior from \autoref{sec:pop_dispersion} is included. The latter forecasted joint constraints are shown in red in \autoref{fig:hypothetical_zigzag_constraints_with_kinematics}. These constraints will have the advantage of not requiring a cosmological prior, and will bring further precision when also combined with the $\zeta_{12}$ measurement from \autoref{sec:zigzag_comparison_to_prior}.
\newpage
\section{Application to previously-studied systems}
\label{sec:literature_application}
%
Although the methods in this paper are novel, constraining $\zeta_{12}$ is convenient in that it comes ``for free'' with a measurement of $\tilde{\eta}$, a cosmological prior and a line-of-sight estimate, without the need for any additional lens modelling. In this section, we will consider two systems for which $\tilde{\eta}$ measurements are available, and use these to derive constraints on $\zeta_{12}$. 
\subsection{Description of systems}
\begin{figure}
    \centering
    \includegraphics[width=0.494\columnwidth]{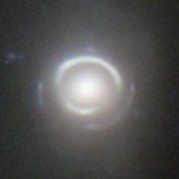}
    \hfill
    \includegraphics[width=0.494\columnwidth]{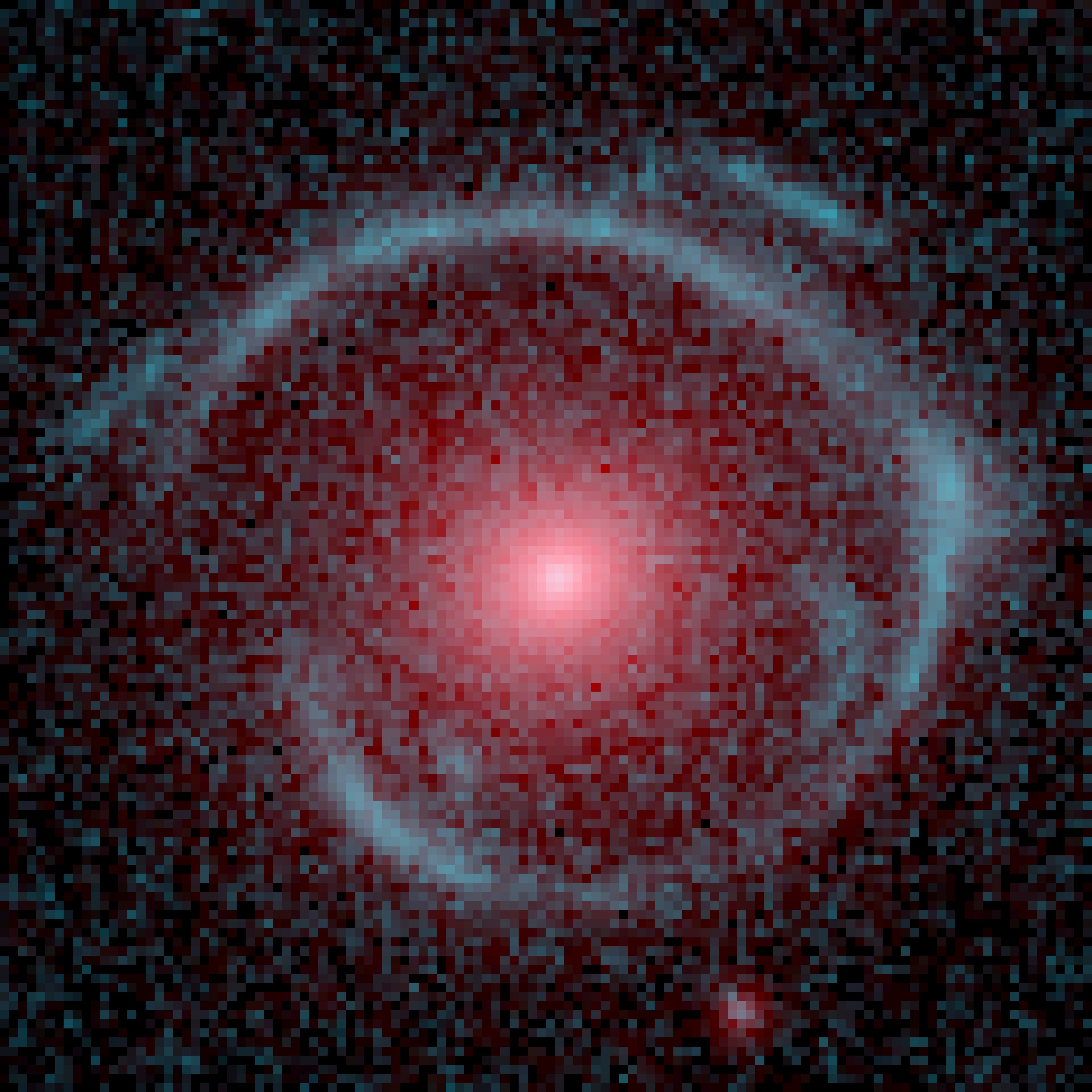}
    \caption{HST images of the DPLs studied in \autoref{sec:literature_application}: the ``Jackpot'' lens J0946 (left) and AGEL0353 (right). AGEL0353 was imaged using HST's Wide Field Camera 3 with the F140W (red) and F200LP (blue) filters by \cite{barone_2025}, the cutout for which is 6$^{\prime\prime}$ in width. \vspace{1em}}
    \label{fig:0946_and_0353_images}
\end{figure}
\subsubsection{J0946}
The `Jackpot' lens J0946+1006, shown in the left panel of \autoref{fig:0946_and_0353_images}, is a massive elliptical galaxy at $z_1=0.222$ \citep{Gavazzi_2008}, which strongly lenses sources at $z_2=0.609$ \citep{Gavazzi_2008} and $z_3=2.035$ \citep{Smith_2022} \citep[with an even more distant third source at $z_4=5.975$,][]{Collett_2020}. After an initial measurement of the cosmological scaling factor in \cite{Collett_2014}, the system was recently revisited in \cite{ballard_2026}, and modelled with elliptical power laws with external shear in both planes. In the case where no velocity dispersion information was used, this study obtained a measurement of $\tilde{\eta}=1.4090^{+0.0035}_{-0.0077}$.
\subsubsection{AGEL0353}
AGEL035346$-$170639 (or DESJ0353-1706), shown in the right panel of \autoref{fig:0946_and_0353_images}, first identified by \cite{Jacobs_2019a,Jacobs_2019} and confirmed by \cite{barone_2025}, consists of a deflector at $z_1=0.617$, a first source/second lens at $z_1=1.460$ and two additional sources at $z_2=1.675$. This system was modelled in \cite{Bowden_2025} with an elliptical power law plus shear for the first deflector and a singular isothermal ellipse for the second, from which a measurement $\tilde{\eta} = 1.0764^{+0.0094}_{-0.0046}$ was obtained.

We note that AGEL150745+052256, like AGEL0353, is a double-plane gravitational lens system observed in the AGEL survey \citep{barone_2025}, from which the cosmological scaling factor has been measured 
\citep{Sahu_2025}. However, this measurement was made with a composite NFW+Chameleon profile, and thus any inferred mass-sheet component of such a profile cannot be interpreted in the same way as for the other systems. Its low $\eta$ value also means that it is less constraining than J0946 or AGEL0353. We therefore omit this system from this analysis.
\newpage
\subsection{Analysis}
\begin{figure}
    \centering
    \includegraphics[width=1\columnwidth]{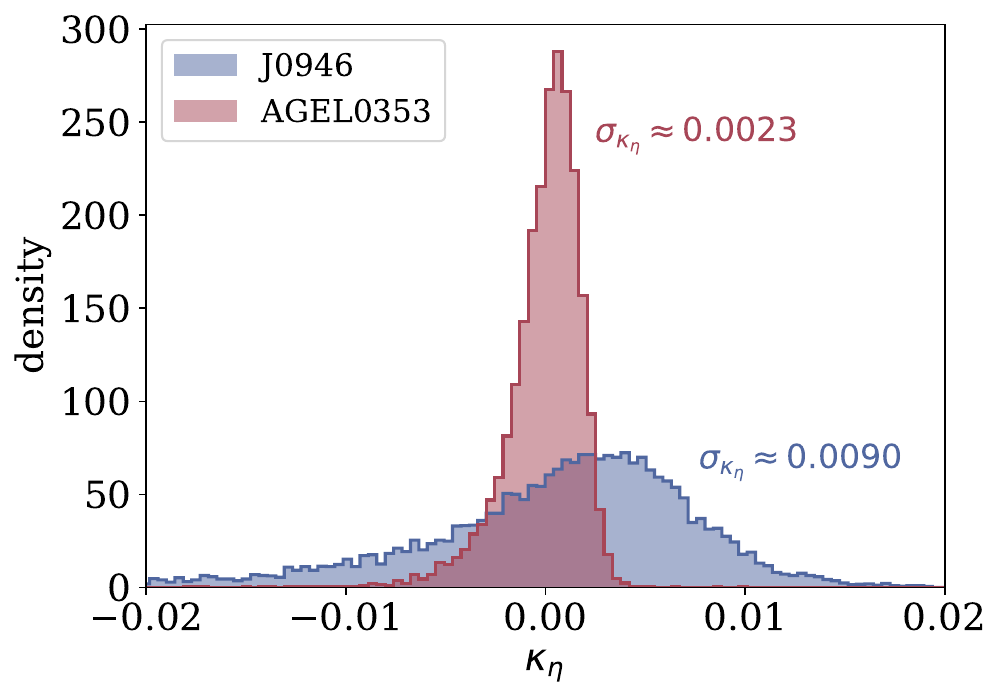}
    \caption{Estimated line-of-sight contribution to the uncertainty in $\eta$, $\kappa_\eta$, for J0946 and AGEL0353, using the methods described in \autoref{sec:zigzag_los_estimate}. The estimate for J0946 is taken from \cite{Johnson_2025}, while that of AGEL0353 was derived in this work. 
    }
    \label{fig:J0946_AGEL0353_los}\vspace{1em}
\end{figure}
Using the same techniques described in \autoref{sec:zigzag_los_estimate}, \cite{Johnson_2025} estimate the uncertainty due to the line of sight to be $\sigma_{\kappa_\eta} = 0.0090$ for J0946. For AGEL0353, we obtain $\sigma_{\kappa_\eta}=0.0023$. The posteriors of these distributions are shown in \autoref{fig:J0946_AGEL0353_los}. As in the case of J1721, these estimates are ``naive'' with respect to the specific line of sight on which the DPLs lie, and simply characterise the distribution of $\kappa_\eta$ values for those redshifts for a random sampling of simulated lines of sight. Once again using the flat $\Lambda$CDM $\Omega\e{m}$ value and uncertainties from \cite{Planck_2020} and combining this with the estimated line of sight contribution, we arrive at a reference value of $\eta=1.386^{+0.0011}_{-0.0071}$ for J0946, and $\eta=1.0911^{+0.0020}_{-0.0014}$ for AGEL0353.
\subsection{Results}
\begin{figure*}
    \centering
    \includegraphics[width=1\textwidth]{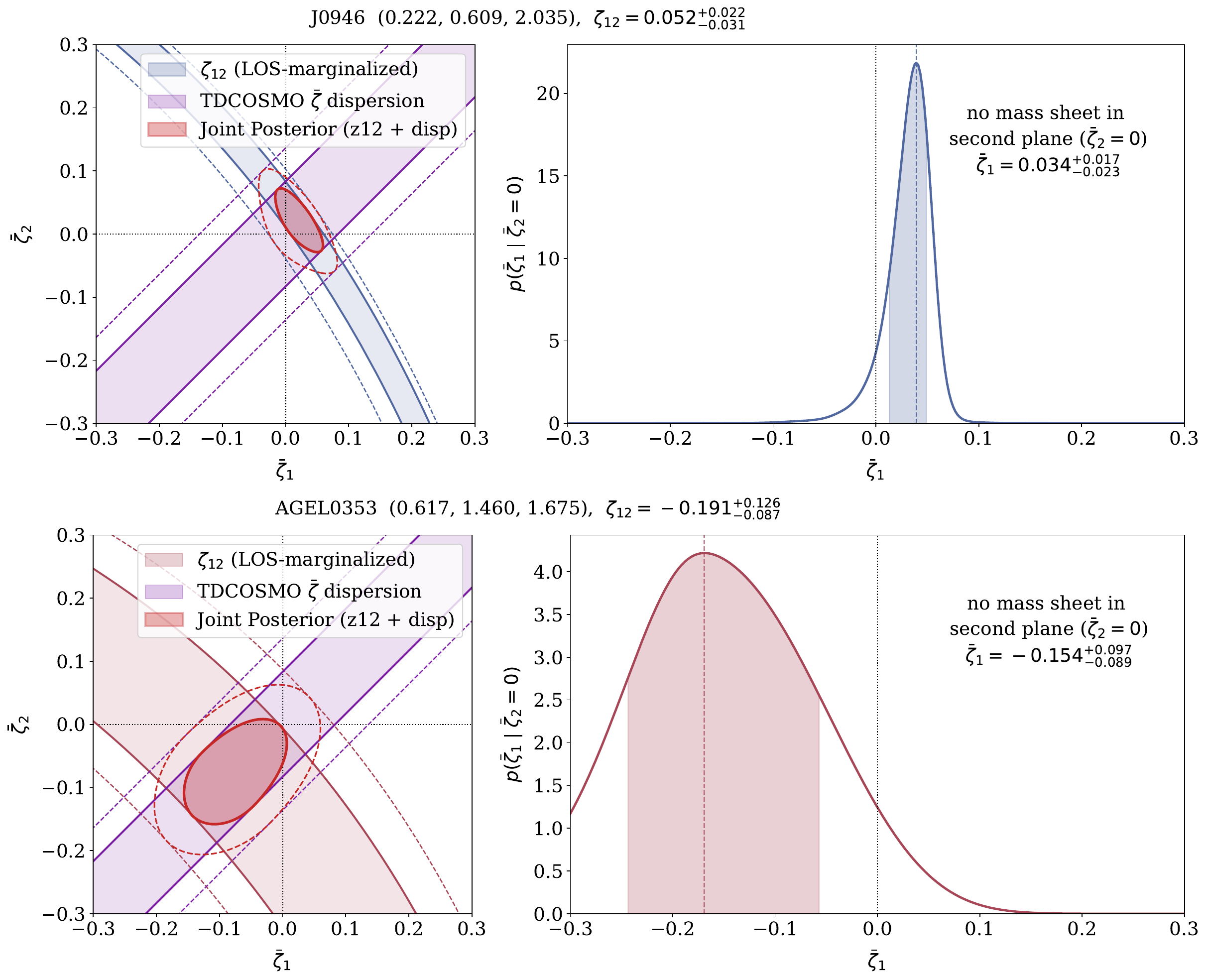}
    \caption{Constraints on the mass-sheet components of J0946 and AGEL0353. In the left column, we show the constraints on $\pa{\bar{\zeta}_1,\bar{\zeta}_2}$ from $\zeta_{12}$, i.e. a comparison between the measured $\tilde{\eta}$ values from \cite{ballard_2026,Bowden_2025} and a reference $\Omega\e{m}$ value from \cite{Planck_2020}, with an additional error budget for the line of sight, in blue. In purple, we show the constraints coming from our measured value of $\sigma_{\bar{\zeta}}$ from \autoref{sec:pop_dispersion}, which assumes that any mass-sheet components in the lenses in \cite{Millon_2020} and the two DPL lens planes are drawn from the same population, described by a Gaussian with scatter $\sigma_{\bar{\zeta}}$. In red, we show the combined constraint. In the panels on the right, we assume that any core component in the second DPL lens planes are too small to act as a mass-sheet for the images we observe, and thus assume $\tilde{\zeta}_2=0$, and plot the constraints implied on $\tilde{\zeta}_1$ from the inferred value of $\zeta_{12}$ only. \vspace{1em}}
    \label{fig:literature_constraints}
\end{figure*}
The larger $\eta$ values of J0946 and AGEL0353 when compared with J1721 imply a higher sensitivity to the cosmological parameters, as well as a larger uncertainty from the line of sight, translating into a larger uncertainty on $\eta$, but also to tighter constraints on $\zeta_{12}$.
\begin{figure}
    \centering
    \includegraphics[width=1\columnwidth]{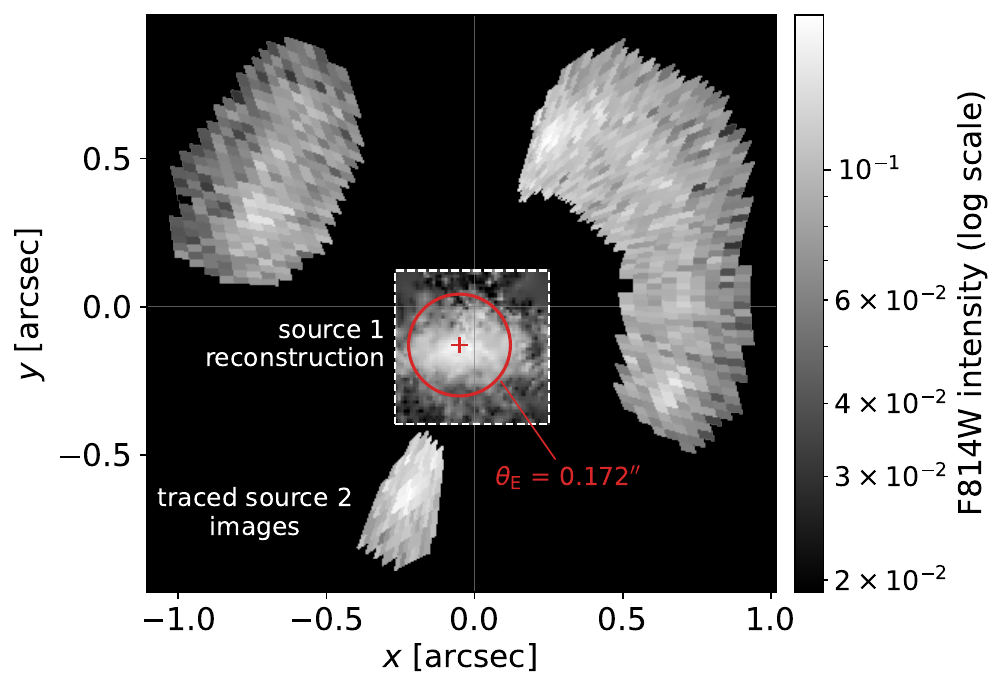}
    \caption{A reconstruction of the first source/second lens in plane 2 of the Jackpot lens system, with the observed light from plane 3 traced back to the positions at which it intersects plane~2. In red, we show the Einstein radius of the second lens mass model inferred in \cite{Li_2026}. \vspace{1em}}
    \label{fig:jackpot_reconstruction}
\end{figure}
\subsubsection{J0946}
Combining the measured $\tilde{\eta}$ value from \cite{ballard_2026} with our reference $\eta$ value, we obtain $\zeta_{12}~=~0.052^{+0.022}_{-0.031}$ for J0946, which mildly disfavours both lenses being power laws. The constraints in the $(\zeta_1, \zeta_2)$ parameter space implied by this measurement are shown in the top panel in \autoref{fig:literature_constraints}. Also plotted on the same axes is the constraint on the scatter in $\zeta$ derived in \autoref{sec:pop_dispersion}, and the combination of the two constraints, which would imply values of $\bar{\zeta}_1 = 0.018^{+0.025}_{-0.027}$, $\bar{\zeta}_2 = 0.018^{+0.034}_{-0.035}$. We again emphasise that this combination rests on strong and likely unreliable assumptions, in particular that the lens galaxies in the TDCOSMO sample in \autoref{sec:pop_dispersion} and the two lens galaxies in J0946 are drawn from the same parent population, and that the dispersion constraints from \autoref{sec:pop_dispersion} effectively marginalise over any dependence on other properties of the galaxy, their redshifts, and so on. 

An alternative choice would be to ignore $\bar{\zeta}_2$ entirely. This is essentially the choice made in \cite{Li_2026}, and implies a value of $\bar{\zeta}_1~=~0.033^{+0.017}_{-0.023}$ without any need for a prior on the lens population dispersion.  

To explore the validity of this approach a bit further, in \autoref{fig:jackpot_reconstruction}, we illustrate the reconstruction of the first source in J0946, as well as the light from the second source as it intersects this plane, en route to forming the image we observe in \autoref{fig:0946_and_0353_images}, as reconstructed from the lens model in \cite{Li_2026}. In red, we plot the Einstein radius from this modelling. This Einstein radius is subject to mismeasurement under a mass-sheet transformation, as
\begin{equation}
    \frac{\tilde{\theta}\e{E}}{\theta\e{E}} = \pa{\frac{\tilde{\eta}-1}{\eta-1}}^\frac{1}{\gamma-1}\pa{1-\zeta_1}^\frac{2-\gamma}{\gamma-1}.
    \label{eq:rescaling}
\end{equation}
A derivation of this can be found in \autoref{sec:power_law_msd_plane_2}. Additionally, the position of the light itself is also rescaled under a mass-sheet transformation, by a factor $\pa{1-\zeta_1}^{-1}$. The ratio of this term to the right hand side of \autoref{eq:rescaling} is thus an additional source of uncertainty in the relevant scales in \autoref{fig:jackpot_reconstruction}, but is nonetheless expected to be small enough to neglect for the purposes of this discussion.

From this figure, we see that the light from source~2 intersects a large region of plane 2, well beyond the outer bounds of the Einstein radius or the light distribution of the second lens, and thus well beyond the scales usually probed by strong lensing. A $\kappa\e{c}$ in \autoref{eq:fid+core} which could maintain a transformation which mimicked an MSD would therefore need to extend far past the characteristic scales on which the mass and baryonic components drop off. Given that galaxies are known to lie within extended dark matter halos, and that such halos can, under the right dark matter model, mimic an MSD of a power law \citep{Blum_2020,Blum_2021}, such a transformation cannot be ruled out. Nonetheless, as these distances grow larger, such a transformation becomes increasingly surprising. When a large population of DPLs is available, it may become possible to probe the scales on which such a transformation may and may not occur, and constrain not only the value of $\zeta$, but also the functional form of $\kappa\e{c}$. This, in turn, may have implications for stellar kinematics modelling \citep[see e.g.][]{Birrer_2020}, and more broadly for our understanding of halo mass profiles at larger radii.

\subsubsection{AGEL0353}
Our results for AGEL0353 are shown in the bottom panel of \autoref{fig:literature_constraints}. With the measured $\tilde{\eta}$ from \cite{Bowden_2025} and our reference $\eta$ value, we infer $\zeta_{12}~=~-0.191^{+0.126}_{-0.087}$ for AGEL0353, which once again mildly disfavours two power law lenses. For the first lens to be a power law, this favours the second having a log-concave profile, and vice versa. 

Proceeding as for J0946, if we combine this constraint with the prior on the scatter in $\tilde{\zeta}$, this implies $\bar{\zeta}_1~=~-0.077^{+0.054}_{-0.052}$, $\bar{\zeta}_2~=~-0.076^{+0.055}_{-0.053}$. As before, this rests on assumptions about a uniformity in the lens population which cannot be extended with certainty to the galaxies in AGEL0353, especially in the second plane. If instead we ignore the mass-sheet in the second plane, we would obtain $\bar{\zeta}_1~=~-0.154^{+0.097}_{-0.089}$, which once again favours a log-concave profile.
%
\section{Forecasts without time delays}
\label{sec:forecasts}
%
\subsection{Which lenses are most constraining?}
\label{sec:good_candidates}
\begin{figure*}
    \centering
    \includegraphics[width=1\textwidth]{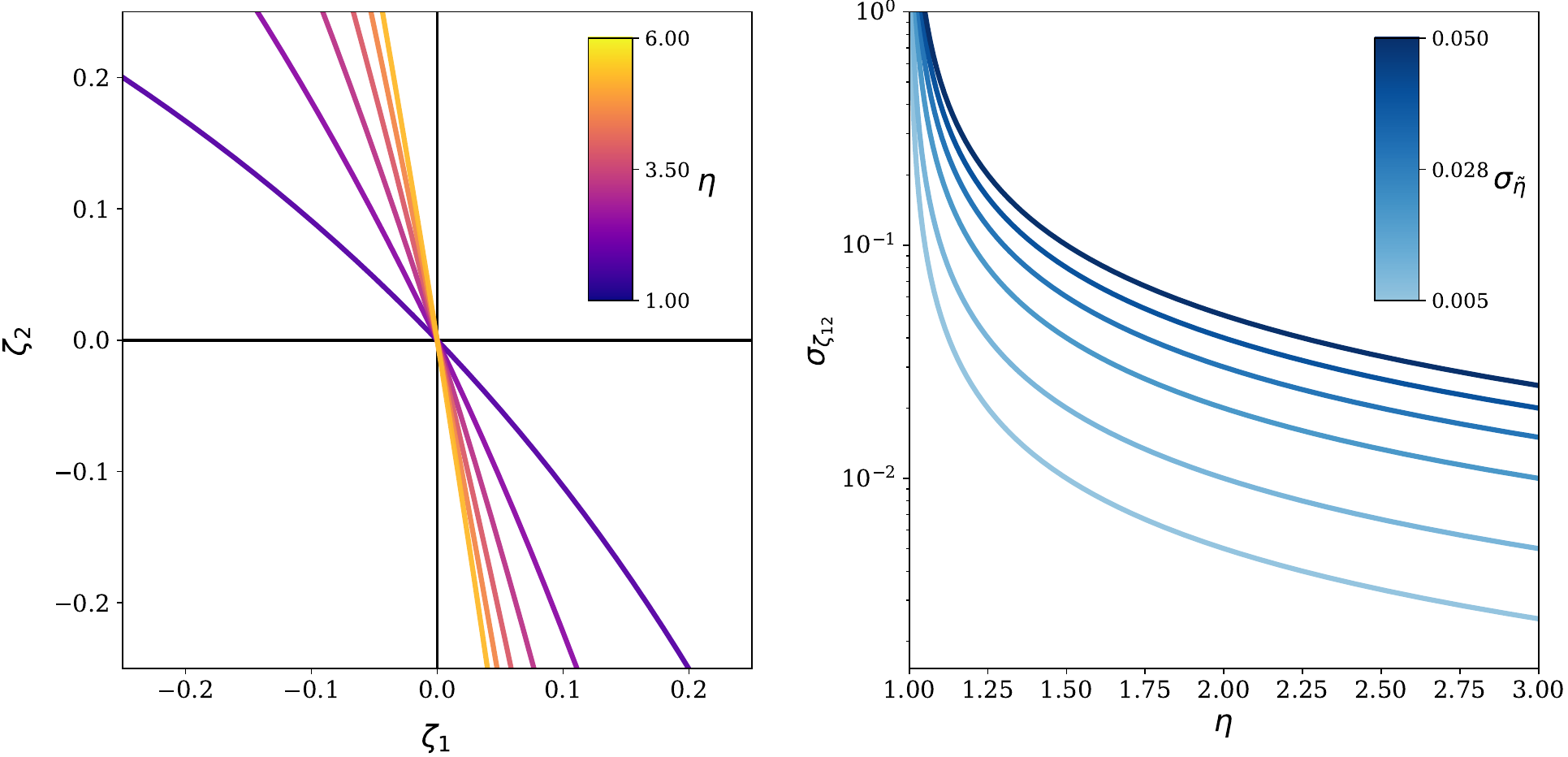}
    \caption{The sensitivity of $\zeta$ constraints to the $\eta$ of a lens system. On the left, we show the slope implied by a $\zeta_{12}$ value, \autoref{eq:zeta_12_definition}, in the $(\zeta_1,\zeta_2)$ parameter space, for different values of the cosmological scaling factor $\eta$. As $\eta$ increases, small changes in $\zeta_1$, the mass sheet in plane 1, correspond to larger and larger changes in $\zeta_2$, the mass sheet in plane 2. On the right, we show the 1$\sigma$ uncertainty in $\zeta_{12}$ as a function of $\eta$, with no contribution from the line of sight. Different values of $\sigma_{\tilde{\eta}}$, the measurement uncertainty in the inferred cosmological scaling factor $\tilde{\eta}$, are plotted as different shades of blue. The best precision on $\zeta_{12}$ comes from systems with higher $\eta$. For a Flat $\Lambda$CDM cosmology, where $\eta$ depends on $\Omega\e{m}$ only, these values are identical irrespective of the lens and source redshift, up to the uncertainty from the line of sight. For a Flat $w_0w_a$CDM cosmology \citep[e.g. from][]{DESI_2024b}, the additional uncertainty in the expansion history at higher redshifts means the best constraints occur when $z_3$ is small at $\eta$ is large. \vspace{1em}}
    \label{fig:theory_eta_dependence}
\end{figure*}
In \autoref{fig:theory_eta_dependence}, we illustrate some of the key relationships which determine which DPLs will be most effective in constraining the mass-sheet degeneracy. In the left panel, we show the relationship between $\eta$ and the corresponding slope of the graph in the $(\zeta_1, \zeta_2)$ parameter space (assuming the measured $\tilde{\eta}$ equals the theoretical $\eta$). What we see from this figure is that, for values of $\eta$ further from 1, we require more and more extreme values of $\zeta_2$ to achieve a given value of $\zeta_1$. Put differently, suppose a value of $\tilde{H}_0$ is measured from a DPL, which differs from your fiducial $H_0$ value. To reconcile these values, $\zeta_1$ must be different from 0. The higher the value of $\eta$ for this system, the more extreme the mass-sheet parameter $\zeta_2$ in plane 2 must be for such a reconciliation to be possible. For a lens like J0946 in which the impact parameter in the second plane is much larger than the Einstein radius of the second lens, this type of transformation becomes increasingly physically bizarre.

In the righthand-side panels, we show the relationship between the value of $\eta$ and the corresponding uncertainty in $\zeta_{12}$, for a range of measurement precisions in $\tilde{\eta}$. From this figure, we see that systems with $\eta\approx 1$ are essentially useless for constraining mass-sheet parameters with this method. As $\eta$ increases, systems quickly grow more constraining, although the precision begins to plateau for larger values of $\eta$. Unsurprisingly, better measurement precision in $\tilde{\eta}$ also corresponds to better constraints on $\zeta_{12}$, and the general increase in line-of-sight uncertainty with redshift means that DPLs with a lower $z_3$ are more precise, an effect compounded when the reference cosmology incorporates dynamical dark energy.

Together, these relationships make the point already hinted at by the results of \autoref{sec:zigzag_application} and \autoref{sec:literature_application}: DPLs with higher values of $\eta$ will be more effective for studying the mass-sheet degeneracy, and well-constrained systems with lower-redshift sources will reach the highest-precision measurements of $\zeta_{12}$.

\subsection{What can we measure with stage-IV surveys?}
\label{sec:surveys}
\begin{figure*}
    \centering
    \includegraphics[width=1\textwidth]{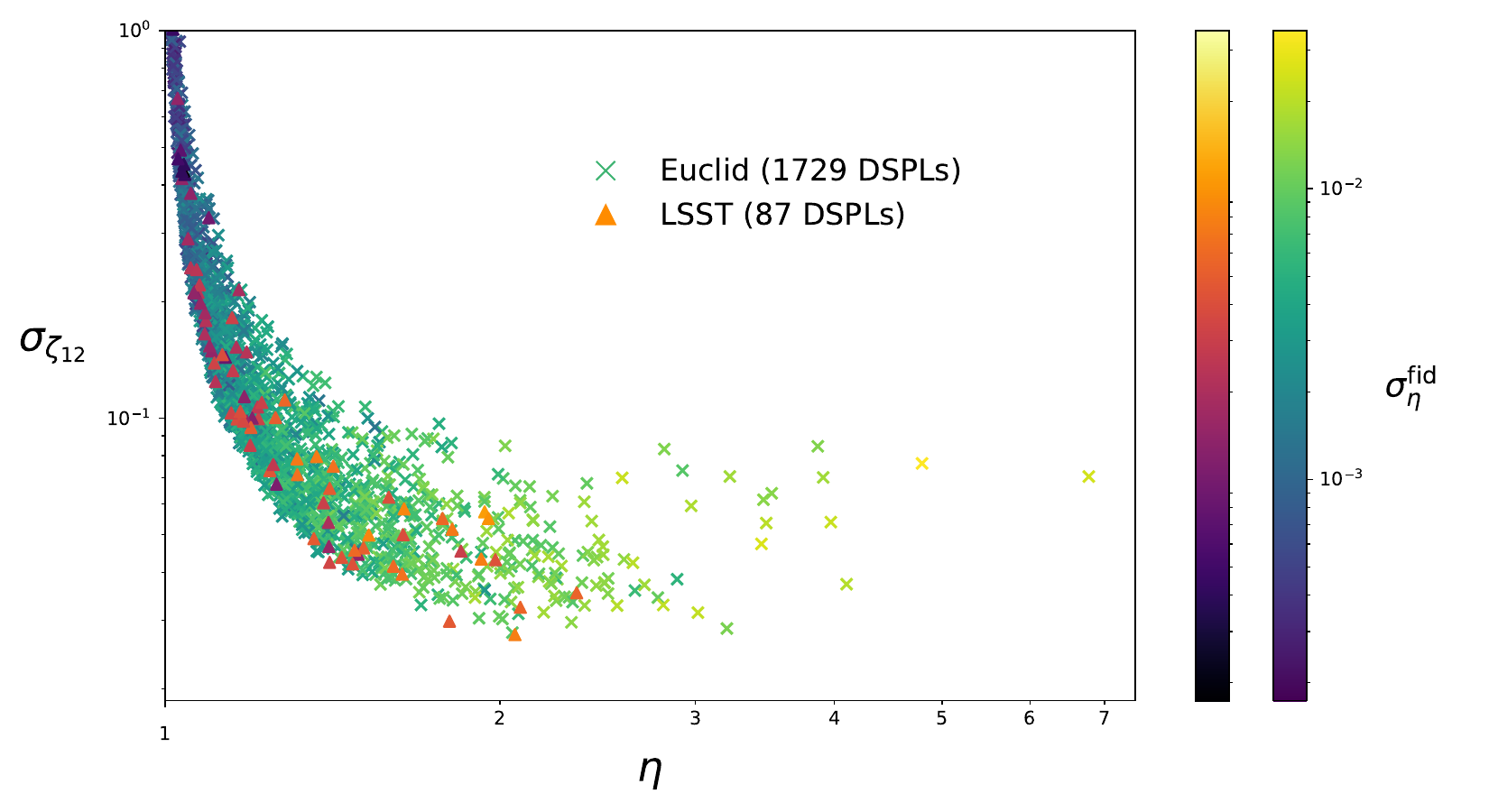}
    \caption{Forecasted $\zeta_{12}$ constraining power of Euclid and LSST double-plane lenses. The vertical axis $\sigma_{\zeta_{12}}$ is the 1$\sigma$ uncertainty of the inferred $\zeta_{12}$ value, plotted against the true cosmological scaling factor $\eta$ on the x-axis. The colours of the points correspond to the uncertainty in the fiducial $\eta$ value, coming both from the uncertainty in cosmological parameters and that from the line of sight. Constraints on $\zeta_{12}$ are obtained by comparing these fiducial values to measured $\tilde{\eta}$ values and their forecasted uncertainties. \vspace{1em}}
    \label{fig:DPL_forecast}
\end{figure*}
The comparisons in \autoref{fig:theory_eta_dependence} are informative, but do not fully capture the relationship between the DPL system redshifts and the expected measurement uncertainty on $\tilde{\eta}$, nor the expected distributions of actual observed systems. In this section, we therefore consider forecasted populations of DPLs from \textit{Euclid} \citep{Euclid_2022} and LSST \citep{LSST_2021}. Both forecasts are generated using the {\sc LensPop} package \citep{Collett_2015}, neglecting mass in the first source, approximating the first plane mass as a singular isothermal ellipsoid (SIE) and modified from the base version to include multiple background sources. Mock Euclid and LSST point spread functions and background noises are used to simulate detectability, and the precision with with which the cosmological scaling factor $\tilde{\eta}$ will be measurable. In the case of LSST DPLs, the sample is limited to a conservative subset predicted as having high resolution imaging and spectroscopic follow-up, and the predicted measurement errors on $\tilde{\eta}$ therefore reflect this assumption. More details of these forecasts can be found in \cite{Sharma_2022}.

In \autoref{fig:DPL_forecast}, we show a forecasted distribution in $\eta$ of 1729 \textit{Euclid} and 87 LSST DPLs, plotted against the forecasted precision with which $\zeta_{12}$ could be constrained from these systems. We also show the precision with which $\tilde{\eta}$ is expected to be measured. The predicted uncertainty in $\zeta_{12}$ accounts for both this forecasted measurement precision, as well as the precision with which the reference $\eta$ value can be predicted. The latter is limited both by the uncertainty in the cosmological parameters as described in \autoref{sec:zigzag_comparison_to_prior}, and from the estimated uncertainty from the line of sight for each system, reported in \cite{Johnson_2025} for the forecasted \textit{Euclid} systems and predicted in this work for the forecasted LSST systems, in both cases using the method described in \autoref{sec:zigzag_los_estimate}. We see that, while the forecasted precision on the predicted $\eta$ values tends to decrease for larger $\eta$ values, as is the case for the measurement precision in $\tilde{\eta}$, it is nonetheless systems with higher $\eta$ values which are more constraining for $\zeta_{12}$. While none of these systems are expected to reach a 1\% precision on $\zeta_{12}$ without follow-up observation for more precise $\tilde{\eta}$ measurements, 
almost 50\% of these systems should still enable $\zeta_{12}$ to be inferred with an uncertainty $<10\%$. 
\begin{figure}
    \centering
    \includegraphics[width=1\columnwidth]{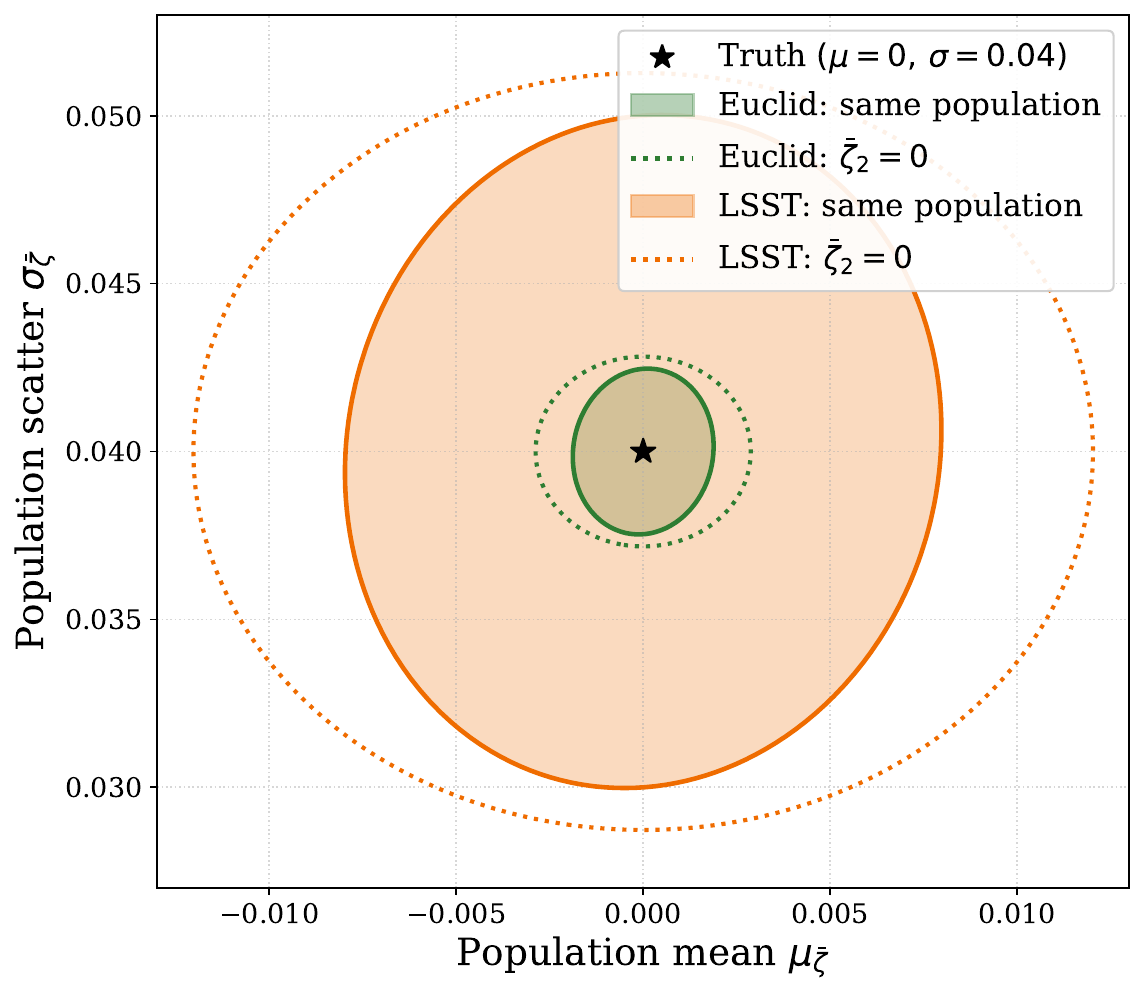}
    \caption{The $1\sigma$ constraining power of a sample of forecasted \textit{Euclid} (in green) and LSST (in orange) DPLs on the mean and intrinsic scatter in the mass sheet parameter $\zeta$, from $\zeta_{12}$ alone. For the solid, filled contours, we assume that the mass-sheet components in the first and second planes of both galaxies are drawn from the same population, described by a Gaussian with mean $\mu_{\tilde{\zeta}}$ and standard deviation $\sigma_{\tilde{\zeta}}$. For the dotted, empty contours, we assume that the mass-sheets in the first lens plane are described by this population, but that the impact of the mass-sheet in the second plane is negligible. \vspace{1em}}
    \label{fig:population_2d_contour_forecast}
\end{figure}

In \autoref{fig:population_2d_contour_forecast}, we illustrate the hypothetical constraining power of these forecasted DPL samples on the mean and scatter of the mass-sheet component. As in \autoref{fig:literature_constraints}, we consider two scenarios,
\begin{enumerate}
    \item \textbf{The ``same population'' assumption :} here, we assume that the mass-sheet components of the lens galaxies in planes 1 and 2 across all forecasted \textit{Euclid} and LSST DPLs are drawn from a single population, described by a Gaussian distribution in $\bar{\zeta}$ with mean $\mu_{\bar{\zeta}}$ and standard deviation $\sigma_{\bar{\zeta}}$.
    \item \textbf{The $\mathbf{\tilde{\zeta}_2=0}$ assumption :} here, we assume that the mass-sheet components of the lens galaxies in plane 1 across all forecasted \textit{Euclid} and LSST DPLs are drawn from a single population, like the one described above, but that the mass-sheet component in plane 2 is negligible on the relevant angular scales.
\end{enumerate}
From \autoref{fig:population_2d_contour_forecast}, we see that, under either assumption, the mean and scatter of this hypothetical population could be constrained extremely precisely, within $\sim 0.01$ for both quantities for LSST, and within $\sim 0.002$ for \textit{Euclid}. In reality, the picture is almost certainly much more complex than either of these scenarios, and in \autoref{sec:discussion} we discuss how this could be approached more rigorously in future analyses with real data. Nonetheless, the above hints at the power such systems may have in understanding the properties of such a population of mass-sheet components, even as the population model increases in complexity. 
%
\section{Forecasts with time delays}
\label{sec:forecasts_time_delays}
%
\subsection{What could we learn from a single time-delay lens with a large $\eta$?}
\begin{figure}
    \centering
    \includegraphics[width=1\columnwidth]{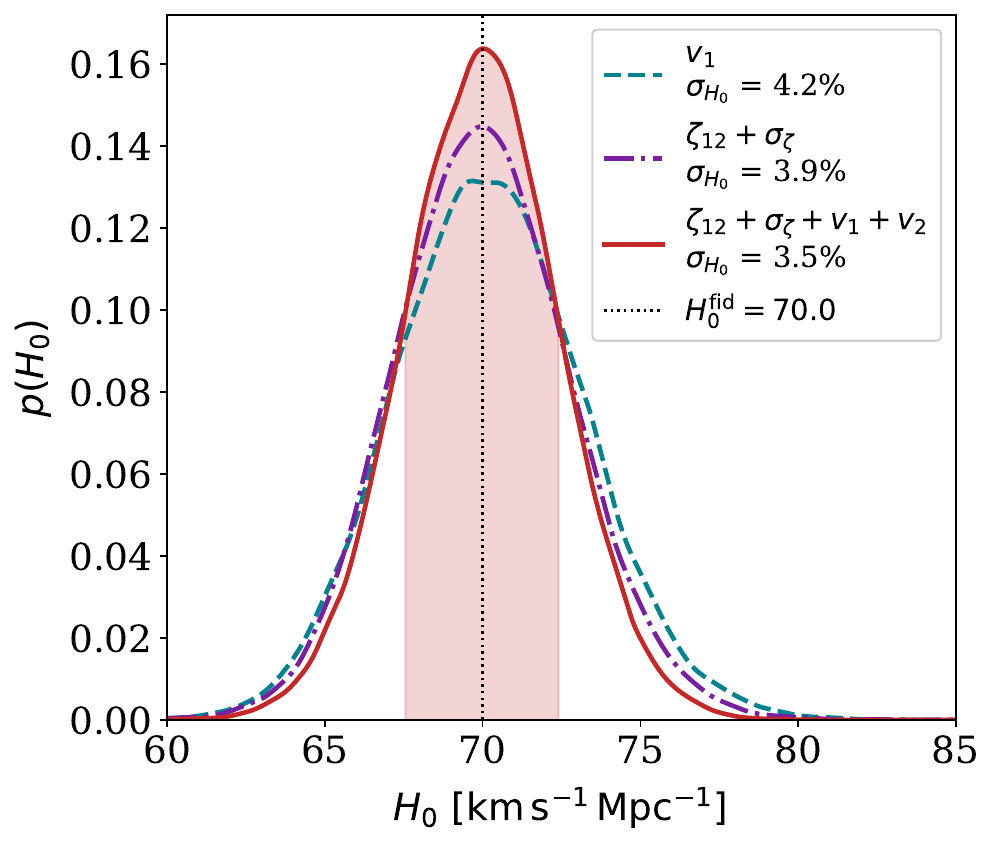}
    \caption{Forecasted $H_0$ constraints with a single ``Jackpot''-like lens system with time delays in one of the source planes. We assume a 2\% contribution each from lens modelling (excluding the mass-sheet) and time delays to this uncertainty. In blue, we imagine a 3\% measurement of $V_1$, in purple we imagine no velocity dispersion measurements, but a $\zeta_{12}$ constraint matching that in \autoref{sec:literature_application} and a $\sigma_{\bar{\zeta}}$ matching that in \autoref{sec:pop_dispersion}. In red, we imagine all of these constraints, plus a 5\% measurement of $V_2$ from plane 2. \vspace{1em}}
    \label{fig:H0_posterior_precision_comparison_zeta2_free}
\end{figure}
J1721, J0946 and AGEL0353 are compelling systems for different reasons. The extreme image configuration of J1721 makes this system highly sensitive to the mass profiles of both lens galaxies, and the quasar in plane 3 means a measurement of $H_0$ is possible, and with it a promising test of cosmology and mass model assumptions. J0946 and AGEL0353, on the other hand, are more sensitive to the relative expansion history and $\zeta_{12}$, but lack compact transient sources from which time delays could be measured. However, given the dramatic increase in the number of DPLs expected in the coming months and years, we may well observe lensed quasars in DPL systems with high $\eta$ values. For such a system, the measurement of $\zeta_{12}$, coupled with a population scatter prior (or the assumption that $\tilde{\zeta}_2=0$) would provide additional constraints on the mass-sheet term with which $H_0$ is degenerate, as would a measurement of $V_2$.

In \autoref{fig:H0_posterior_precision_comparison_zeta2_free}, we imagine a ``Jackpot''-like system, with identical redshifts and measurement precision on $\tilde{\eta}$, but for which we can also measure time delays in one of the two source planes. We assume that the uncertainties in these time delays contribute a 2\% uncertainty on $H_0$, the lens modelling (ignoring the MSD) an additional 2\%, and that $\kappa\h{LOS}_{12}$ is constrained from weighted number counts to the expected $\sim 1\%$ for these redshifts. In blue, we show the constraints possible with a 3\% uncertainty in $V_1$ to constrain the internal mass-sheet component. In purple, we show the constraints with no velocity dispersion measurement, but with a measurement of $\tilde{\eta}$, a reference $\eta$ value from \cite{Planck_2020} and the dispersion constraint from \autoref{sec:pop_dispersion}. In red, we show the combined constraints possible with all of these constraints, plus a 5\% measurement of $V_2$ from the velocity dispersion in the images of the second lens.

As discussed in \autoref{sec:what_we_learn_time_delays}, when comparing the $\sim 3\%$ $H_0$ measurement of such a system before accounting for the internal mass-sheet to a fiducial $H_0$ value, a value of $\zeta_1$ is implied. Any tension between this value and the parameter space allowed by other constraints (from $\tilde{\eta}$ measurements, a prior on the expansion history, measurements of $V_1$ and $V_2$, and a prior on the population or population scatter in $\zeta$), would point to a flaw in one of these assumptions. While such a tension is unlikely to reach more than 2-3$\sigma$ for a single system, the more such systems are studied, the more pressure is put on any deviant constraint and the assumptions beneath it.
\subsection{How much does the second plane improve our $H_0$ constraints?}
\begin{figure*}
    \centering
    \includegraphics[width=1\textwidth]{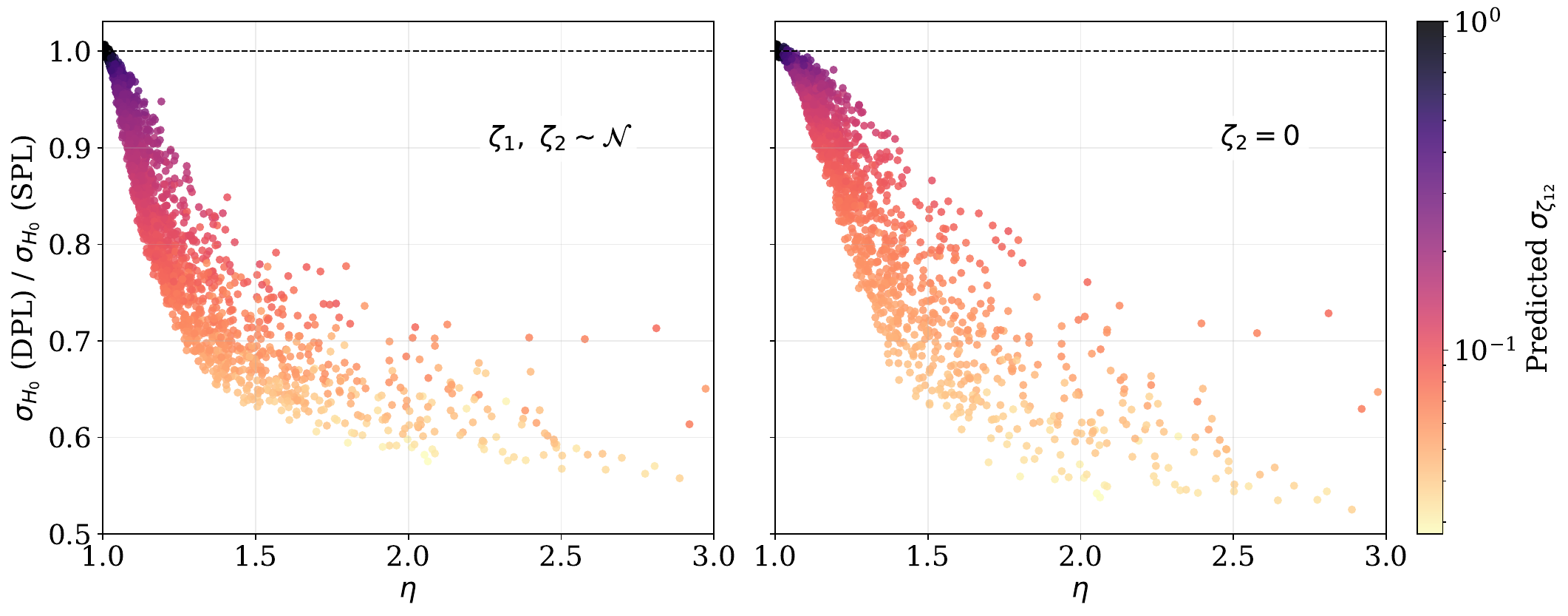}
    \caption{The benefit of the second plane in hypothetical \textit{Euclid} and LSST-like double-plane time-delay lenses, under the two simple scenarios considered in \autoref{sec:surveys}. On the vertical axis, we plot the ratio of the uncertainty $H_0$ precision when constraints derived from the second source are added to that when they are not, with the cosmological scaling factor on the horizontal axis. The colours of the points correspond to the forecasted precision with which $\sigma_{\zeta_{12}}$ could be constrained. The slight scatter around 1 for $\eta\approx1$ is simply a result of sampling noise in our forecast. \vspace{1em}}
    \label{fig:euclid_lsst_fractional_H0_precision_change}
\end{figure*}
Let us finally ask how much the precision in the Hubble constant is generally improved when a second source plane is present. To that end, we once again consider the population of \textit{Euclid} and LSST lenses from \autoref{sec:surveys}, but this time supposing that these systems have measurable time delays for one of the two source planes. These forecasts were not generated with compact sources in mind; a majority of them will not have measurable time delays, and so we will not derive any forecasted population-based constraints. Nonetheless, a subset of these systems will have quasars or supernovae within their source galaxies, and considering these populations gives a flavour of the impact of the second source plane for realistic combinations of parameters and measurement precisions.

In \autoref{fig:euclid_lsst_fractional_H0_precision_change}, we plot the improvement that comes with having access to the second source plane. We begin by assuming a measurement uncertainty in $\pa{1-\kappa\h{LOS}_{12}}^{-1}H_0$ of $\sim 4\%$, a combination of a 2\% uncertainty from time delays, a 2\% uncertainty from the lens model, and a 3\% uncertainty in $V_1$. The baseline $H_0$ uncertainty is then estimated by adding the estimated error budget from the line-of-sight convergence, giving us $\sigma_{H_0}$ (SPL), the 1$\sigma$ uncertainty from constraints which require only single-plane lensing. To this, we then add the additional constraints coming from the predicted $\zeta_{12}$ measurement. We once again consider the two simple scenarios from \autoref{sec:surveys}, taking the same uncertainty in the $\zeta$ population scatter from \autoref{sec:pop_dispersion} in the first scenario. The combination of these constraints gives $\sigma_{H_0}$ (DPL), the 1$\sigma$ uncertainty when adding $\zeta_{12}$ information that is only available in DPLs. In the figure, we show the ratio of these quantities.

We see that, in both scenarios, for low values of $\eta$, there is little improvement, but when $\eta$ is larger, the gain in precision with our mock values approaches 50\%. When a 5\% $V_2$ measurement is added to the above constraints, we expect improvements of 20\% - 50\% in $H_0$ for all \textit{Euclid} or LSST-like DPLs with time delays, even when $\eta\sim 1$.

While the population prior we took from \autoref{sec:pop_dispersion} is again likely oversimplified, it nonetheless serves as a placeholder for any similar prior on the scatter in $\zeta$ for these galaxies, and could come from other DPLs, as in \autoref{sec:surveys}, or velocity dispersion measurements of other lenses or even similar non-lens galaxies \citep[see e.g.][]{Birrer_2020}.

\section{Conclusion and discussion}
\label{sec:discussion}
%
In this article, we have argued that, while they do not truly break the mass-sheet degeneracy, double-plane lenses are nonetheless sensitive to changes in the internal mass-sheet component $\bar{\zeta}$ through their measured cosmological scaling factors. By comparing this scaling factor to a fiducial relative expansion history, with no need for an assumed value of $H_0$, we showed in \autoref{sec:mass-sheet constraints} that we infer $\zeta_{12}$, a combination of the mass-sheet components in the two lens planes. With a line-of-sight convergence estimate, coupled with a prior on either the ``similarity'' of lens galaxies, in other words the permitted strength of the variance in $\bar{\zeta}$ values across different lenses, or alternatively a stellar kinematics measurement in the first plane, the degeneracy is fully broken, and we can estimate the value of $\bar{\zeta}$ in both planes. A similar combination of mass-sheet terms is constrained by a measurement of the scaling factor and the velocity dispersion in plane 2, without the need for a reference cosmology.

In \autoref{sec:zigzag_application}, we considered the J1721, the ``Einstein zig-zag lens''. With a simple elliptical power law modelling of the two lens planes, considering only the image positions of the multiply lensed quasar and fixing the perturber models to their values from \cite{Schmidt2025}, we measure $\tilde{\eta} = 1.0166 ^{+0.0011}_{-0.0011}$. With a prior on $\Omega\e{m}$ from \cite{Planck_2020} and a ``naive'' line-of-sight convergence estimate, we infer $\zeta_{12} = -0.22 ^{+0.12}_{-0.13}$. Upcoming velocity dispersion data and extended source modelling will further refine this measurement.

In \autoref{sec:literature_application}, we considered J0946, the ``Jackpot'' lens, and AGEL0353. For these systems, taking $\tilde{\eta}$ measurements from \cite{ballard_2026,Bowden_2025}, naive line-of-sight convergence estimates from \cite{Johnson_2025} and this work, and the same \cite{Planck_2020} $\Omega\e{m}$ reference, we infer $\zeta_{12}~=~0.052^{+0.022}_{-0.031}$ and $\zeta_{12}=-0.191^{+0.126}_{-0.087}$ respectively. The result for J0946 disfavours both lens planes having strongly log-concave (or ``steepening'') mass profiles, while those from J1721 and AGEL0353 disfavour both their respective lens planes having strongly log-convex (or ``flattening'') mass profiles. In all three cases, the $\zeta_{12}$ results alone very mildly ($\sim1-2\sigma$) disfavour no-mass sheet transform in either plane (i.e. both planes being power laws). 

When combined with a population prior on the scatter in mass-sheet parameter for the lens planes, derived from the scatter in $H_0$ from \cite{Millon_2020}, we infer values of $\bar{\zeta}_1$ and $\bar{\zeta}_2$, the individual mass-sheet components of each plane. If instead we assume that the mass-sheet in the second planes are negligible, our constraint of $\zeta_{12}$ alone would imply values of $\bar{\zeta}_1$, without the need for a population prior. Neither of these assumptions is robust -- the selection function for the time-delay lenses in \cite{Millon_2020} is completely different to the criteria by which galaxies can constitute a DPL like J0946, especially in the case of the second lens plane. On the other hand, fixing $\bar{\zeta}_2=0$ without a well-motivated uncertainty is similarly overconfident, and we do not yet understand galaxy morphologies well enough to robustly make this choice. Nonetheless, these results are an illustrative first step, and we discuss why the latter assumption might have merit, at least for certain systems. Finally, we emphasise that our LOS convergence estimates for the three DPLs are agnostic to the specific over or under density on their individual lines of sight, thus follow-up study with weighted number counts from wide-field observations will be essential to obtaining truly robust results.

While constraints from individual systems are interesting in their own right, and allow us to test the consistency between the assumptions beneath them, the true power of DPL MSD constraints will come from population studies. In \autoref{sec:good_candidates}, we show that constraints on $\tilde{\zeta}_{12}$ are most compelling when $\eta$ is large, and the relative contribution of each plane to $\tilde{\zeta}_{12}$ changes with $\eta$. LSST and \textit{Euclid} are expected to discover $\mathcal{O}(100)$ and $\mathcal{O}(1000)$ DPLs respectively, across a wide range of $\eta$ values, from which we expect to constrain $\zeta_{12}$ within 10\% for hundreds of systems. In \autoref{fig:population_2d_contour_forecast}, we show that this population will enable a highly precise inference on the population of mass-sheet components in lens galaxies, under the assumption that this population is universal and Gaussian. Such an assumption is clearly too strong, but from these tentative results, it does not seem too optimistic to expect that a \textit{Euclid} or LSST-like population could constrain a more flexible and realistic parameterisation of the mass-sheet population. 

Our current knowledge of galaxy mass profiles is too limited for us to confidently rule out the kinds of physical structures which could mimic a mass sheet in a lens galaxy. Nonetheless, whatever such structures may be present, it should be uncontroversial to expect that they are correlated with other properties of these galaxies, such as the density of their environments, their redshift, or the Einstein and half-light radii. Furthermore, the extent to which such structures could mimic a true mass-sheet transformation almost certainly depends on the physical extent of these structures, compared to the point at which the lensed light intersects the lens plane. We anticipate that, with a large sample of DPLs, a population description of $\bar{\zeta}$ which accounts for these other observables will be able to reveal such correlations, in particular when combined with time delays or velocity dispersion measurements from single plane systems (and perhaps also a subset of DPLs). Eventually, these population studies will allow us to place tight priors on the MSD whenever a new system is analysed.

Our parameterisation of a mass-sheet-transformed mass profile, as in the majority of studies, is motivated by the strict mathematical degeneracy from \cite{Falco_1985,Schneider_2013} and others. While transparent in its effect on the Hubble constant, it has the disadvantage of not teaching us much that is useful about the profiles of galaxies themselves. As shown in e.g. \cite{Li_2026}, when the mass-sheet degeneracy is broken, the visible and dark components of the lens galaxy can be disentangled. As our understanding of the MSD improves, so too will our modelling of galaxies, and with it our understanding of the nature of dark matter \citep[see][]{Blum_2020, Blum_2021}. 

Finally, in \autoref{sec:forecasts_time_delays}, we consider the prospects for measuring $H_0$ from DPLs, when one of the two sources has measurable time delays. By adding a $\zeta_{12}$ measurement and a prior on the scatter in $\zeta$ (or the assumption that $\zeta_2=0$), we expect the uncertainty in $H_0$ to be almost halved for systems with $\eta\gtrsim 1.5$, compared to analogous single plane lenses where $\zeta$ is constrained with stellar kinematics only. When the velocity dispersion in plane 2 is also measured, we expect improvements between 15\% and 50\% across all systems, regardless of their $\eta$. 

While tentative, these results emphasise the promise of DPLs in helping to answer some of the most important questions in physics today. The measured distance ratios, velocity dispersions and time delays of such systems are each sensitive to mass profiles of lens galaxies, the Hubble constant and the relative expansion history of the universe. As more and more of these systems are observed and studied, these effects can be untwined, and with this, increasingly precise tests of the Hubble tension, the nature of dark matter and the evolution of dark energy are made possible.

\section{Author contributions and acknowledgments}

\paragraph{Author contributions}
DPJ led the project, contributed to the initial conceptualisation, carried out the analysis in sections 4, 5.3, 5.4, 5.5, 6, 7, 8 and the appendices, and wrote the initial draft.
MM  contributed to the initial conceptualisation and modelled the lensed quasar J1721+8842.
PF contributed to the initial conceptualisation, made \autoref{fig:double-plane_lensing}, and reviewed and edited the draft.
GQ contributed to the initial conceptualisation, carried out the estimation of astrometric uncertainty in section 5.6, and reviewed and edited the draft. 
TL made \autoref{fig:jackpot_reconstruction} and contributed to the discussion.
WE contributed through discussion of the concepts, review and editing of the draft.
DJB modelled J0946, provided the $\eta$ value that enabled the analysis in Section 6 and 8.1, and reviewed and edited the draft.
DB contributed to section 6, and reviewed and made minor edits to the draft.
TC provided the forecasted Euclid and LSST DPL samples, and contributed to the discussion.
DG contributed to section 5.6, and reviewed and edited the draft.

\paragraph{Acknowledgments}
DPJ acknowledges support by the FirstRand Foundation, South Africa, and the Centre National de la Recherche Scientifique of France. 
MM acknowledges support by the SNSF (Swiss National Science Foundation) through return CH grant P5R5PT\_225598 and Ambizione grant PZ00P2\_223738. 
PF acknowledges support from the French \emph{Agence Nationale de la Recherche} through the ELROND project (ANR-23-CE31-0002).
WE acknowledges the support by the Italian Ministry of University and Research (grant FIS2023-01611, CUP C53C25000300001).
DJB acknowledges the support of the Australian Research Council Discovery Project DP230101775.
This work has received funding from the European Research Council (ERC) under the European Union's Horizon 2020 research and innovation program (LensEra: grant agreement No 945536). TC is funded by the Royal Society through a University Research Fellowship.
This work used the Python packages emcee \citep{Foreman-Mackey_2013}, ChainConsumer \citep{Hinton_2016}, and Lenstronomy \citep{Birrer_2021b}.

\newpage

\begin{appendix}

\section{A mass-sheet transformed power law in plane 2}
\label{sec:power_law_msd_plane_2}

Suppose we have a lens mass profile which takes the form given in \autoref{eq:fid+core}, with a $\kappa\e{mod}$ given by \autoref{eq:EPL_convergence}. Here, we show that a mass-sheet transformation in the second plane of a DPL system enables such a lens to be modelled with a power law only. Applying \autoref{eq:mst_2_k_core} to this profile, and as usual dropping the $\kappa\e{c}$, we would obtain
\begin{align}
    \tilde{\kappa}\h{lens}_2(\tilde{\btheta}_2) &= \nu_3\lambda^{-1}\kappa\h{EPL}_2\pa{\lambda^{-1}\tilde{\btheta}_2} \nonumber \\ &\quad + \nu_3\lambda^{-1}\zeta_2 + \tilde{\eta}\pa{1-r_\eta\nu_3\lambda^{-1}}.
\end{align}
As before, the constant piece on the right hand side of this equation is the mass-sheet component we infer, $\tilde{\zeta}_2$, which can always be transformed to 0 with the appropriate choice of $\tilde{\eta}$. 

As for the transformed power-law component of this model, firstly note that, since $\tilde{\btheta}_2 = \lambda\btheta$ simply multiplies the vector quantity $\btheta$ by a constant, we can introduce $\tilde{\vartheta}=\lambda\vartheta$ as the mass-sheet transformed elliptical radius. When the convergence profile in the second plane is evaluated at $\lambda^{-1}\btheta$, we can likewise write the argument of $\kappa\h{EPL}_2$ as $\lambda^{-1}\tilde{\vartheta}$, and, with some straightforward algebra, write
\begin{equation}
    \nu_3\lambda^{-1}\kappa\h{EPL}_2\pa{\lambda^{-1}\tilde{\vartheta}_2}
    = \frac{(3-\gamma)}{2}\pa{\frac{\nu_3^\frac{1}{\gamma-1}\lambda^\frac{\gamma-2}{\gamma-1}\theta\e{E}}{\tilde{\vartheta}}}^{\gamma-1}. 
\end{equation}
From this expression, we confirm that, provided $\tilde{\eta}$ is left free to vary, we always have the flexibility to model plane 2 with an elliptical power law. The value of $\tilde{\eta}$ will compensate any mass-sheet component $\zeta_2$ according to \autoref{eq:zeta_12_definition}, and the Einstein radius of the power law model will absorb the rescaling of both the convergence profile and its argument. What we would interpret as the Einstein radius from our model in the second plane, $\tilde{\theta}\e{E}$, would relate to the true parameter $\theta\e{E}$ via 
\begin{equation}
    \frac{\tilde{\theta}\e{E}}{\theta\e{E}} = \nu_3^\frac{1}{\gamma-1}\lambda^\frac{\gamma-2}{\gamma-1}.
\end{equation}
For an isothermal lens model with $\gamma=2$, the difference between the modelled $\tilde{\theta}\e{E}$ and the true $\theta\e{E}$ is entirely determined by the difference between the modelled $\tilde{\eta}$ and the true $\eta$.

\section{$\zeta_{12}$ in the $\beta$ formalism}
\label{sec:beta_formalism}
%
\subsection{The lens equation}

In \autoref{sec:theory}, we adopted the convention to define lensing quantities (displacement angles, potentials etc) with respect to the subsequent lens or source plane, such that the displacement angle in the first plane must be scaled by $\eta$ in \autoref{eq:eff_lens_2}. We could instead choose to define all quantities with respect to the final source plane of the multi-plane system in question. Under this choice, the definition of $\balpha_2(\btheta_2)$ from \autoref{eq:alpha2_los} remains unchanged, but $\bet{\balpha_1}(\btheta)$ from \autoref{eq:alpha1_los} becomes
\begin{equation}
    \bet{\balpha_1}(\btheta) \equiv \frac{\rd\bet{\psi_1}}{\rd\btheta} = \frac{1-\kappa_{13}}{1-\kappa_3}\bet{\bar{\balpha}_1}[(1-\kappa_1)\btheta], \label{eq:alpha1_los_beta} 
\end{equation}
where $\bet{\bar{\balpha}_1}(\btheta)$ is the corresponding displacement angle of the first lens only in this new formalism, $\bet{\psi_1}(\btheta)$ the lens potential, and we use the colour to differentiate \textcolor{Maroon}{quantities redefined within this formalism}, versus quantities consistent with their definition from before.

With this convention, the lens equations can be written
\begin{gather}
    \btheta_2 = \btheta - \beta\bet{\balpha_1}(\btheta),
    \label{eq:lens1} \\
    \btheta_3 = \btheta - \bet{\balpha_1}(\btheta
    )-\bar{\balpha}_2(\btheta_2),
    \label{eq:lens2}    
\end{gather}
where $\beta$ is the cosmological scaling factor, given by
\begin{equation}
    \beta = \eta^{-1} = \frac{\pa{1-\kappa_{12}}\pa{1-\kappa_3}}{\pa{1-\kappa_2}\pa{1-\kappa_{13}}}\frac{D_{12}D_3}{D_2D_{13}}. \label{eq:barbeta}
\end{equation}
\subsection{Time delays}
In the $\beta$ formalism, $\tau_{12}$, $\tau_{23}$ and $\phi_{23}$ in \autoref{eq:time_delays} retain their original definitions. $\phi_{23}$ remains physically unchanged, but can be rewritten in terms of the $\bet{\psi_1}$ defined in the $\beta$ convention via
\begin{equation}
    \phi_{12}(\btheta,\btheta_2) = \frac{1}{2}(\btheta-\btheta_2)^2-\beta \bet{\psi_1}(\btheta), \label{eq:phi_1_beta} 
\end{equation}
In \cite{Johnson_2026}, we showed that
\begin{equation}
    \frac{\tau_{23}}{\tau_{12}} = \frac{\beta}{1-\beta},
\end{equation}
which gives us the flexibility to express \autoref{eq:time_delays} as either
\begin{equation}
    T_{123} = \pa{1-\kappa\h{LOS}_{12}}\bar{\tau}_{12}\pa{\phi_{12}+\frac{\beta}{1-\beta}\phi_{23}}, \label{eq:tau12_factored_beta} 
\end{equation}
or
\begin{equation}
    T_{123}  = \pa{1-\kappa\h{LOS}_{23}}\bar{\tau}_{23}\pa{\frac{1-\beta}{\beta}\phi_{12}+\phi_{23}},    \label{eq:tau23_factored_beta}
\end{equation}
where we've abbreviated $T(\btheta, \btheta_2, \btheta_3)$ as $T_{123}$. 
\subsection{The mass sheet degeneracy}
Once again, the mass sheet degeneracy manifests as an uncertainty in both $\lambda$ and $\tilde{\beta}$, the measured cosmological scaling factor in the $\beta$ convention. The effect of this degeneracy on $\balpha_2(\btheta_2)$, $\kappa_2(\btheta_2)$ and $\psi_2(\btheta_2)$ is exactly equivalent to the expressions given in Eqs.~\ref{eq:mst_2_k} and \ref{eq:full_psi_MSD_2}, while in the first plane,
\begin{gather}
    \bet{\tilde{\kappa}_1}(\btheta) = \lambda\bet{\kappa_1}(\btheta)+\frac{1-r_\beta\lambda}{\tilde{\beta}}, \\
    \tilde{\kappa}_2(\tilde{\btheta}_2)  = \frac{\nu_3}{r_\beta\lambda}\kappa_2\pa{r_\beta^{-1}\lambda^{-1}\tilde{\btheta}_2}+\frac{1-\lambda^{-1}\nu_3}{\tilde{\beta}},
\end{gather}
and the corresponding potentials are 
\begin{align}
    \bet{\tilde{\psi}_1}(\btheta) &= \lambda\bet{\psi_1}(\btheta)+\frac{1}{2}\frac{1-r_\beta\lambda}{\tilde{\beta}}\btheta^2, \label{eq:full_psi_MSD_1_tilde_beta}\\
    \tilde{\psi}_2(\tilde{\btheta}_2) &= \lambda r_\beta\nu_3\psi_2\pa{r_\beta^{-1}\lambda^{-1}\tilde{\btheta}_2} \nonumber\\ &\hspace{2cm} +\frac{1}{2}\frac{1-\lambda^{-1}\nu_3}{\tilde{\beta}}\tilde{\btheta}^2_2, \label{eq:full_psi_MSD_2_tilde_beta}
\end{align}
where we have defined
\begin{equation}  
    \nu_3 = \frac{\beta}{\tilde{\beta}}\frac{1-\tilde{\beta}}{1-\beta}, \qquad 
    r_\beta = \frac{\tilde{\beta}}{\beta}.
\end{equation}
From \cite{Johnson_2026}, the mass-sheet transformed time delays can then be written as
\begin{equation}
\tilde{T}_{123}
    = \frac{\tilde{\tau}_{12}}{\tau_{12}}r_\beta\lambda T_{123}.
\end{equation}
\subsection{What do double-plane lenses constrain?}
If we once again assume that the true mass profiles of our lens galaxies are given by \autoref{eq:fid+core}, drop the $\kappa\e{c}(\btheta)$ which has no impact on the scales we're interested in, and assume that both planes are modelled without the $\zeta$ components, this implies
\begin{gather}
    \lambda\bet{\zeta_1}+\frac{1-r_\beta\lambda}{\tilde{\beta}} = 0, \label{eq:beta_implication_1} \\
    \frac{\nu_3}{r_\beta\lambda}\zeta_2+\frac{1-\lambda^{-1}\nu_3}{\tilde{\beta}} = 0, \label{eq:beta_implication_2}
\end{gather}
where $\bet{\zeta_1}$ is the mass-sheet component of plane 1 in the $\beta$ convention, which relates to the original parameter as $\bet{\zeta_1} = \eta\zeta_1$.

With a mass-sheet free modelling of the lens planes, the parameters $\lambda$ and $\tilde{\beta}$ are once again constrained in their relationship to the physical quantities $\zeta_1$, $\zeta_2$ and $\beta$ as Eqs.~\eqref{eq:beta_implication_1} and \eqref{eq:beta_implication_2}. Rearranging, this implies
\begin{equation}
    \bet{\zeta_1}+\zeta_2-\beta\bet{\zeta_1}\zeta_2 = \frac{1}{\beta}\pa{\frac{\beta-\tilde{\beta}}{1-\tilde{\beta}}}=\bet{\zeta_{12}}, \label{eq:zeta_12_beta_convention}
\end{equation}
Using the relationship between $\bet{\zeta_1}$ and $\zeta_1$, we see that $\bet{\zeta_{12}} = \zeta_{12}$, and so, from \autoref{eq:zeta_12_beta_convention}, we confirm that the measured cosmological scaling factors in the $\beta$ convention, $\tilde{\beta}$, and in the $\eta$ convention, $\tilde{\eta}$, follow the expected relationship $\tilde{\eta}=\tilde{\beta}^{-1}$. We can use this fact when interpreting constraints on $\tilde{\beta}$ reported in the literature.
%
\section{Estimating the dispersion in $\zeta$ from uncorrected $H_0$ measurements}
\label{sec:dispersion_estimate}
%
In the following, we go through the details of the method we use to put an upper bound on $\sigma_{\bar{\zeta}}$, the intrinsic scatter in the mass-sheet component of lens galaxies, and apply this method to a sample of lenses from \cite{Millon_2020}.
%
\subsection{Introducing the notation and models}
%
We will express measurements of $\tilde{H}_0$ from a time-delay lens $i$ as 
\begin{equation}
    m^{H_0}_i = \frac{H_0}{\left(1-\kappa\h{LOS}_i\right)(1-\zeta_i)}+\epsilon_i. \label{eq:measurement_model}
\end{equation}
where the measurement error $\epsilon_i$, as well as $\kappa\h{LOS}_i$ and $\zeta_i$ are specific to lens $i$. The reported measurement $m^{H_0}_i$ and the line-of-sight convergence $\kappa\h{LOS}_i$ are represented by probability distributions
\begin{equation}
    p_i\pa{m^{H_0}_i}, \qquad p_i\pa{\kappa\h{LOS}_i},
\end{equation}
which we will approximate as asymmetric Gaussian (split normal) distributions, defined by the reported measurements/estimates and their upper and lower 1$\sigma$ uncertainties.

We will assume that the internal mass-sheet parameters $\zeta_i$ of the $N$ lenses are drawn from a common population with mean $\mu_{\bar{\zeta}}$ and dispersion $\sigma_{\bar{\zeta}}$,
\begin{equation}
    \zeta_i \sim \mathcal{N}\pa{\mu_{\bar{\zeta}},\sigma^2_{\bar{\zeta}}}. \label{eq:zeta_distribution}
\end{equation}
Because $H_0$ and $\bar{\zeta}$ are degenerate within the observations we consider here, we define the quantities
\begin{gather}
    \mu \equiv \frac{H_0}{1-\bar{\zeta}}, \label{eq:mu_definition} \\
    f_{\bar{\zeta}} \equiv \frac{\sigma_{\bar{\zeta}}}{1-\bar{\zeta}}. \label{eq:f_zeta_definition}
\end{gather}
The quantity $\mu$ is what is measured by a population of time-delay lenses where the impact of the $\kappa\h{LOS}_i$ are corrected for, but $\mu_{\bar{\zeta}}$ remains unknown. $f_{\bar{\zeta}}$ represents the fractional population scatter in $1-\bar{\zeta}$. For each lens, we also define the dimensionless variable 
\begin{equation}
    x_i \equiv \frac{1-\bar{\zeta}_i}{1-\bar{\zeta}}, \label{eq:x_definition}
\end{equation}
meaning, following from Eqs.~\eqref{eq:zeta_distribution}, \ref{eq:f_zeta_definition} and \ref{eq:x_definition}, we can write
\begin{equation}
    x_i \sim \mathcal{N}\pa{1,f^2_{\bar{\zeta}}}, \label{eq:x_pop}
\end{equation}
and rewrite our measurement model from \autoref{eq:measurement_model} as
\begin{equation}
    m^{H_0}_i = \frac{\mu}{\pa{1-\kappa\h{LOS}_i}x_i}+\epsilon_i. \label{eq:new_measurement_model}
\end{equation}

\subsection{Estimating the fractional scatter in $\bar{\zeta}$}

\begin{table*}[t]
\centering
\label{tab:TDCOSMO_sample}
\begin{tabular}{lcccccc}
\toprule
Lens & RXJ1131 & HE0435 & J1206 & WFI2033 & PG1115 & DES0408 \\
\midrule
$\tilde{H}_0$ & $76.1^{+ 3.6}_{-4.3}$ & $71.7^{+5.1}_{-4.6}$ & $67^{+5.7}_{-4.8}$ & $72.6^{+3.3}_{-3.5}$ & $83^{+7.8}_{-7.0}$ & $74.6^{+2.5}_{-2.9}$   \\
\bottomrule
\end{tabular}
\caption{Values and upper and lower uncertainties for $\tilde{H}_0$ with power-law fits, taken from the left panel in Fig. 6 in \cite{Millon_2020}, used to estimate the dispersion in $\zeta$. 
}
\end{table*} 

For a proposed $f_{\bar{\zeta}}$, a realisation of the population is generated by drawing a value of $x_i$ for each, according to the distribution in \autoref{eq:x_pop} defined by that $f_{\bar{\zeta}}$ value, together with draws of
\begin{equation}
    m^{H_0}_i \sim p_i\pa{m^{H_0}_i}, \qquad \kappa_i \sim p_i\pa{\kappa\h{LOS}_i}.
\end{equation}
The sampled values of $x_i$, $m^{H_0}_i$ and $\kappa_i$, in turn, define the value of $\mu$ implied by each lens, according to \autoref{eq:new_measurement_model}. Thus, each realisation produces a set of $\mu_i$ values corresponding to the $N$ lenses. For the correct value of $f_{\bar{\zeta}}$, the lens-to-lens variation in $\mu_i$ should be explained by the scatter implied by this value, compatible with a common underlying $\mu$. 

Our goal is to construct and maximise the likelihood of the measured $H_0$ values for each realisation of $\pa{\mu,f_{\bar{\zeta}}}$, by marginalising over the quantities $\kappa_i$ (using an estimation from simulations, possibly weighted by number counts) and $x_i$ (drawn according to the given value of $f_{\bar{\zeta}}$). We could write
\begin{multline}
    p_i\pa{m^{H_0}_i | \mu, f_{\bar{\zeta}}}  = \int p_i\pa{\kappa_i}p\pa{x_i|f_{\bar{\zeta}}} \\ \times p_i\pac{m^{H_0}_i| \frac{\mu}{\pa{1-\kappa_i}x_i}} \rd \kappa_i \rd x_i,
\end{multline}
where $p\pa{x_i|f_{\bar{\zeta}}}$ is the Gaussian distribution we have assumed in \autoref{eq:x_pop}.

Treating the lenses independently, we can write the total likelihood as
\begin{equation}
    p\pa{\mathrm{data}|\mu,f_{\bar{\zeta}}} = \prod^N_{i=1}p_i\pa{m^{H_0}_i|\mu,f_{\bar{\zeta}}}.
\end{equation}
To obtain our prior on the fractional scatter, we must marginalise this over the nuisance parameter $\mu$,
\begin{equation}
    p\pa{f_{\bar{\zeta}} | \mathrm{data}} \propto p(f_{\bar{\zeta}})\in p \pa{\mathrm{data}|\mu,f_{\bar{\zeta}}}p\pa{\mu}\rd \mu.
\end{equation}
This is done numerically, via Monte Carlo methods, rather than analytically.

\subsection{Estimating the absolute scatter}
\label{sec:absolute_scatter}

Using the relationships given in \autoref{eq:mu_definition} and \ref{eq:f_zeta_definition}, we can write the absolute scatter in $\bar{\zeta}$, $\sigma_{\bar{\zeta}}$, as 
\begin{equation}
    \sigma_{\bar{\zeta}} = f_{\bar{\zeta}} \frac{H_0}{\mu}.
\end{equation}
Thus, to obtain this estimate, we would need to use a prior on the true value of $H_0$, with higher $H_0$ values corresponding to higher estimates of $\sigma_{\bar{\zeta}}$. Then, sampling this prior in conjunction with our distribution of $\paac{\mu,f_{\bar{\zeta}}}$ from the previous step, we can obtain a posterior on $\sigma_{\bar{\zeta}}$.

\subsection{Application to the TDCOSMO sample}
\label{sec:tdcosmo_scatter_application}

We apply the procedure described above to the $H_0$ measurements from \cite{Millon_2020}, shown in \autoref{tab:TDCOSMO_sample}, for which the impact of the line-of-sight convergence has already been incorporated into the estimate and its uncertainty, but for which there is no correction for the internal mass-sheet. The procedure is therefore simplified, as the marginalisation over $\kappa\h{LOS}_i$ has effectively already been carried out. Our method requires a prior on $\mu$ (\autoref{eq:mu_definition}), for which we use the value of $74.2 ^{+1.6}_{-1.6}$\ksmpc derived for these systems in \cite{Millon_2020}.

\begin{figure}[h]
    \centering
    \includegraphics[width=1\columnwidth]{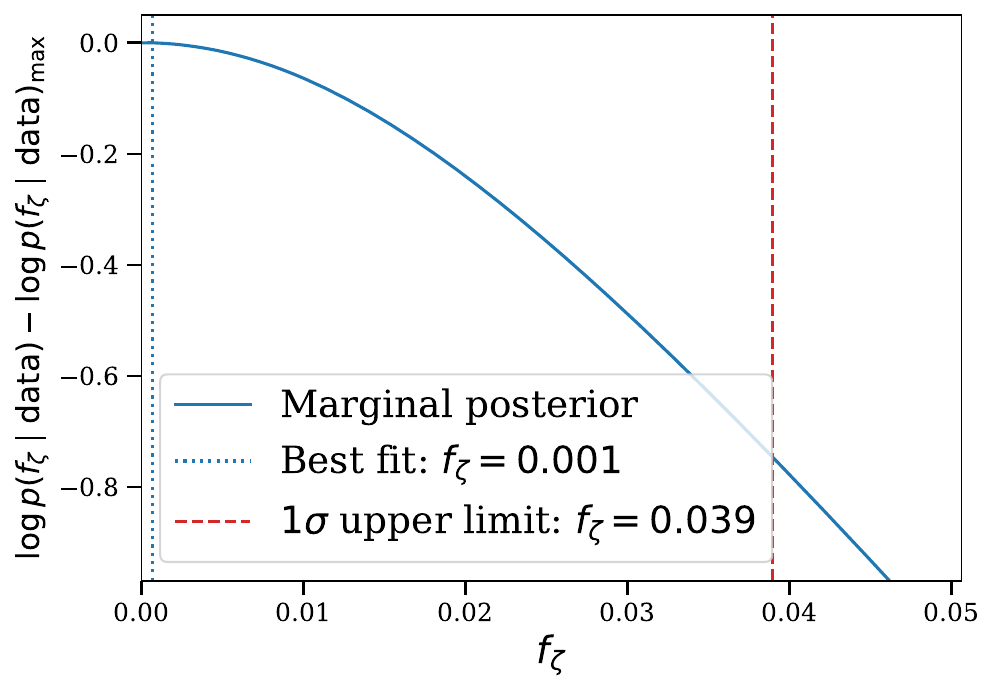}
    \caption{The posterior for the fractional scatter in $\zeta$ for the sample of lenses in \autoref{tab:TDCOSMO_sample} from \cite{Millon_2020}. In red, we show the 1$\sigma$ upper bound we can place on this scatter. \vspace{1em}}
    \label{fig:posterior_fractional_scatter}
\end{figure}

With these values, the data do not show evidence for scatter in $\bar{\zeta}$, and we can place a $1\sigma$ upper bound on $f_{\bar{\zeta}}$ of 0.039, as shown in \autoref{fig:posterior_fractional_scatter}. As discussed in \autoref{sec:absolute_scatter}, converting this to a constraint on the absolute scatter requires a prior on $H_0$, with larger values of $H_0$ implying a larger scatter. The constraint on $f_{\bar{\zeta}}$ could be incorporated directly into an inference of $H_0$ from a DPL with time delays without such a prior, but for the illustrative purposes of this work, we conservatively marginalise our result over the $H_0 = 73.04 \pm 1.04$ constraint from \cite{Riess_2022}, and obtain a $1\sigma$ upper bound on $\sigma_{\bar{\zeta}}$ of 0.039.


This result should not be interpreted as a reliable indicator that there is no scatter in $\bar{\zeta}$, but rather that the current measurement uncertainties in $H_0/\pa{1-\zeta}$ values are too high and the sample of values too small for the detection of any such scatter. Any meaningful scatter which was present would likely become detectable in a larger sample with precise measurements of $H_0/\pa{1-\zeta}$. 

\end{appendix}

\bibliographystyle{apsrev4-1}
\bibliography{main}

\end{document}